\documentclass[journal]{./IEEEtran}

\ifCLASSINFOpdf
  \usepackage[pdftex]{graphicx}
  \graphicspath{{./fig/}}
  \DeclareGraphicsExtensions{.pdf,.jpeg,.png}
\else
  \usepackage[dvips]{graphicx}
  \graphicspath{{./fig/}}
  \DeclareGraphicsExtensions{.eps}
  \fi

\usepackage[utf8]{inputenc} 
\usepackage[english]{babel} 
\usepackage{scrhack} 
\usepackage[scr=euler]{mathalfa} 
\usepackage{amssymb}
\usepackage{amsmath}
\usepackage{algorithm, algpseudocode}
\usepackage{trsym}
\usepackage{color}

\usepackage{emptypage}

\usepackage{pdflscape}
\usepackage{psfrag}
\usepackage{pstool}

\usepackage{subcaption}
\usepackage{mathtools}

\graphicspath{{./fig/}}
\DeclareGraphicsExtensions{.eps}
\usepackage{pgfplots}
\pgfplotsset{compat=newest}
\pgfplotsset{plot coordinates/math parser=false}
\usepackage{tikz}
\tikzset{>=latex}
\pgfsetxvec{\pgfpoint{1mm}{0}}
\pgfsetyvec{\pgfpoint{0}{1mm}}

\usetikzlibrary{patterns,plotmarks,positioning,calc,arrows.meta}
\usepackage{epstopdf}
\usepackage{accents}
 \allowdisplaybreaks

\usepackage{bm, upgreek}
\newcommand*{\lambdafHl}{{\mf{\Lambda}_h^{(l)}}}
\newcommand*{\pil}{p}
\newcommand*{\textpl}{pl}
\newcommand*{\datasscript}{d}
\newcommand*{\upmuvec}{{\bm{\upmu}}}

\newcommand{\ve}{\mathbf}
\newcommand{\m}{\mathbf}
\newcommand{\veel}[1]{#1} 
\newcommand*{\mel}[1]{\left[#1\right]}

\newcommand{\mf}[1]{\mathbf{\tilde{\mathbf{#1}}}} 
\newcommand{\vef}[1]{\mathbf{\tilde{\mathbf{#1}}}} 

\newcommand{\est}[1]{\hat{#1}}

\newcommand*{\E}[1]{\mathrm{E}\left\{ #1 \right\}}
\newcommand*{\diag}[1]{\mathrm{diag}\left( #1 \right)}
\newcommand*{\tr}[1]{\mathrm{tr}\left( #1 \right)}

\newcommand*{\reallabel}[0]{\mathrm{Re}}
\newcommand*{\imaglabel}[0]{\mathrm{Im}}

\newcommand*{\e}{\mathrm{e}}

\newcommand*{\lmmse}{\text{LMMSE}}

\newcommand*{\cfo}{\text{CFO}}

\newcommand*{\ici}{\text{ICI}}

\newcommand*{\cp}{\text{cp}}
\newcommand*{\cpofdm}{\text{cp}}

\newcommand*{\delayspread}{\tau_\text{RMS}}
\newcommand*{\corrupt}{\text{cor}}

\newcommand*{\downs}{{pl}}

\newcommand*{\noisef}{\ve{v}}
\newcommand*{\varnoisef}{\sigma_v^2}

\newcommand*{\noisefpil}{\ve{v}''}  

\newcommand*{\mElemNoisePil}{\veel{v}''[m]}  

\newcommand*{\mElemNNoisePil}{\veel{v}'''[m]}  

\newcommand*{\noiseffull}{\ve{v}'}  

\newcommand*{\varphil}{{\varphi_l}}
\newcommand*{\varphilest}{\est{\varphi}_l}

\newcommand*{\lambdaNfl}{\mf{\Lambda}'^{(l)}}
\newcommand*{\lambdafl}{{\mf{\Lambda}^{(l)}}}

\newcommand*{\Yfl}{\vef{y}^{(l)}}

\newcommand*{\Yfrxl}{\vef{y}_r^{(l)}}

\newcommand*{\Yfdownl}{\vef{y}_{\downs}^{(l)}}

\newcommand*{\xtl}{\ve{x}'^{(l)}}  
\newcommand*{\xt}{\ve{x}'}  
\newcommand*{\Xfl}{\vef{x}'^{(l)}}  

\newcommand*{\datal}{\ve{d}^{(l)}}
\newcommand*{\mElemdatal}{\veel{d}^{(l)}[m]}
\newcommand*{\kElemdatal}{\veel{d}^{(l)}[k]}

\newcommand*{\kElemdatacorruptl}{\veel{d}_\corrupt^{(l)}[k]}

\newcommand*{\dataestl}{\est{\ve{d}}^{(l)}}

\newcommand*{\kElemdataestlperfect}{\veel{\est{d}}^{(l)}_{\text{perf}}[k]}
\newcommand*{\kElemdataestlnonperfect}{\veel{\est{d}}^{(l)}_{\text{non-p}}[k]}
\newcommand*{\kElemdataestl}{\veel{\est{d}}^{(l)}[k]}

\newcommand*{\kElemdatacfol}{\veel{d}_\cfo^{(l)}[k]}

\newcommand*{\dataicil}{\ve{d}_\ici^{(l)}}

\newcommand*{\mElemdataicil}{\veel{d}_\ici^{(l)}[m]}

\newcommand*{\piloticil}{\ve{p}_\ici^{(l)}}
\newcommand*{\pestl}{\est{\ve{p}}^{(l)}}

\newcommand*{\mElempestl}{\est{\veel{p}}^{(l)}[m]}

\newcommand*{\mElempicil}{\veel{p}^{(l)}_\ici[m]}

\newcommand*{\mElemp}{\veel{p}[m]}

\newcommand*{\mElemwp}{\veel{w}_p[m]}

\newcommand*{\errl}{\ve{e}^{(l)}}

\newcommand*{\kElemnoisellr}{\veel{w}}

\begin{document}

\title{Analysis and Compensation of Carrier Frequency Offset Impairments in Unique Word OFDM}

\author{Christian Hofbauer,~\IEEEmembership{Member,~IEEE}, Werner
  Haselmayr,~\IEEEmembership{Member,~IEEE}, Hans-Peter
  Bernhard,~\IEEEmembership{Senior Member,~IEEE}, and Mario~Huemer,~\IEEEmembership{Senior Member,~IEEE}
\thanks{The authors are with the Silicon Austria Labs GmbH, Linz, Austria
  (christian.hofbauer@silicon-austria.com, hans-peter.bernhard@silicon-austria.com), with the
  Institute of Signal Processing,
Johannes Kepler University Linz, Linz, Austria (mario.huemer@jku.at), and with the Institute for Communications Engineering and
RF-Systems, Johannes Kepler University Linz, Linz, Austria (werner.haselmayr@jku.at), respectively.

This work has been supported by Silicon Austria Labs (SAL), owned by the
Republic of Austria, the Styrian Business Promotion Agency (SFG), the federal
state of Carinthia, the Upper Austrian Research (UAR), and the Austrian
Association for the Electric and Electronics Industry (FEEI).}
}


\maketitle

\begin{abstract}
Unique Word-orthogonal frequency division multiplexing (UW-OFDM) is known to
provide various performance benefits over conventional cyclic
prefix (CP) based OFDM. Most important, UW-OFDM features excellent spectral sidelobe
suppression properties and an outstanding bit error ratio (BER) performance. Carrier
frequency offset (CFO) induced impairments denote a challenging task for OFDM
systems of any kind. In this work we thoroughly investigate the CFO effects on UW-OFDM and
compare it to conventional OFDM. Different
CFO compensation approaches with different computational complexity are
considered throughout this work, assessed against each other, and the residual
CFO error of these approaches is analyzed by deriving analytical error models. A mean squared error analysis carried out after data estimation reveals a significantly higher
robustness of UW-OFDM over CP-OFDM against CFO effects. Additionally, the
conducted BER simulations generally support this conclusion for various
scenarios, ranging from uncoded to coded transmission in a frequency
selective environment.

\end{abstract}


\IEEEpeerreviewmaketitle

\section{Introduction}
In Unique Word (UW)-OFDM, introduced in \cite{Huemer10_1}, the conventional
cyclic prefix (CP) in the guard interval is replaced  by a deterministic
sequence -- the UW. The
introduction of the UW within the interval of the discrete Fourier transform
(DFT) entails the introduction of redundancy in the frequency domain, which can beneficially be utilized to obtain superior spectral shaping characteristics
\cite{Huemer12_1,Rajabzadeh13,Rajabzadeh14,Rajabzadeh18} or outstanding bit error ratio (BER)
performance for linear \cite{Huemer11_1,Huemer12_1,Hofbauer16_1,Qasem2021,Chema2016}, non-linear
\cite{Huemer12_2,Onic14,Onic13,Steendam16}, iterative\footnote{The term \emph{iterative receiver} refers to an iterative
  exchange of realibility information between detector and decoder
  \cite{Douillard95, Tuechler02_1}.} \cite{Haselmayr14,Haselmayr19}
as well as neural network (NN)
\cite{Baumgartner2023,Bognar2021} based receivers. 
 
Various other approaches known as PRP-OFDM (pseudorandom prefix) \cite{Muck06}, KSP-OFDM (known symbol padding)
\cite{Welden08}, OFDM with a PN (pseudo noise) sequence \cite{Tang07}, TDS-OFDM
(time domain synchronous) \cite{China06,Ong10,Tang07}, or even OFDM with Unique
Word \cite{Jingyi02} implement deterministic sequences in the guard
interval. Sharing with UW-OFDM the common idea of a deterministic sequence in
the guard interval, only UW-OFDM implements it inside the DFT, and thus solely
benefits from the introduced redundancy and the resulting beneficial properties.

So far, investigations of UW-OFDM have primarily focused on principle performance
bounds and were based on various idealized assumptions, such as perfect timing,
carrier phase or carrier frequency synchronization. Investigations aside from
idealized scenarios have been limited to the impact of imperfect channel
estimation on the BER behavior \cite{Huemer12_1}, peak to average power ratio
(PAPR) and peak to minimum power ratio (PMR) analyses
\cite{Rajabzadeh2021,Huber12_1,Rettelbach12}, as well as feasibility considerations in terms of computational complexity \cite{Huemer11_1,Onic11}.

A major challenge in OFDM arises from a carrier frequency offset (CFO), which makes accurate estimation and compensation of
this offset essential. This task is often divided into an acquisition phase providing an
initial rough correction of the entire packet based on the preamble
\cite{IEEE99}, and a subsequent tracking phase for a finer correction on a
symbol-by-symbol basis. In this paper, we only address the tracking phase.


Existing work on UW based systems together with CFO considerations focused on the
time domain based estimation thereof based on UWs \cite{Imec00,
  Witschnig03,Aboltins2012, Kim2010, Ehsa202004}, mainly in context of UW based
single-carrier systems. In \cite{Hofbauer20,Hofbauer20_1} we extended
considerations to UW based OFDM systems with frequency pilot tone assisted CFO estimation.


In this work, all CFO related estimation tasks rely on pilot frequency tones as
presented in \cite{Hofbauer20,Hofbauer20_1} rather than on UWs. Therefore, the same estimation method
can be applied to all considered systems, i.e., also to non-UW based
reference systems, thus ensuring a fair comparison among them. As such, an
investigation of UWs for estimation purposes is not within the scope of this
work. UWs in this work are limited to evaluations on how they might
degrade the system performance as a result of their erroneous handling
in the receiver due to CFO. For all investigations in this work, we assume a
CFO to stay constant during transmission as well as a perfect timing
synchronization enabling perfect packet detection.

In this work, we aim at analyzing the impact of a CFO on the performance of UW-OFDM based systems, whereas the main contributions can be summarized as follows\footnote{We note that
 this work is based on unpublished parts of the PhD thesis in
 \cite{Hofbauer16}.}. We elaborate on various CFO
compensation approaches and assess their performance w.r.t.\ computational complexity and CFO compensation
capabilities. We derive analytical models describing the residual error
after CFO compensation, which can further be utilized as a reliability information in a
communication system. Finally, we conduct BER simulations of a whole
communication chain in different variants (with different combinations of coding rate and modulation alphabet)
in order to assess the  performance also on a system level. Throughout all
evaluations, classical CP-OFDM serves as a reference.  

We note that this work expands its conference version \cite{Hofbauer20_2} in
various aspects. While the main focus in \cite{Hofbauer20_2} was laid on the
analysis of the different CFO effects on various UW-OFDM setups, this work rather
focuses on handling those in the context of a whole UW-OFDM based transceiver system. Besides conventional
common phase error (CPE) correction already addressed in \cite{Hofbauer20_2},
we additionally consider more sophisticated CFO compensation methods that
take into account all CFO induced impairments. We develop
computational complexity reduced variants of thereof and compare their
performance against the original implementation. Moreover, we extend
performance assessment from mean squared error (MSE)
considerations in \cite{Hofbauer20_2} to evaluations at system level by means
of BER simulations. For this, we additionally develop accurate analytical error models
for the different CFO compensation methods and utilize them to provide
reliability information to the channel decoder at the receiver.

The remainder of this work is organized as follows. We start with a recap
of the UW-OFDM signaling model incorporating a CFO in Sec.~\ref{sec:review}. In
Sec.~\ref{sec:impact_cfo_mse}, we investigate the remaining impact of
these effects on the UW-OFDM performance by means of MSE analyses after applying different CFO
compensation approaches. Subsequently, in Sec.~\ref{sec:error_model} we derive analytical models for the residual CFO
error remaining from the
different approaches presented in Sec.~\ref{sec:impact_cfo_mse}. Next, we
evaluate the CFO impact on the overall system performance in terms of BER simulations in
Sec.~\ref{sec:impact_cfo_ber}, and we finally conclude our work in Sec.~\ref{sec:conclusion}.

\smallskip
\textit{Notation:} We use lower case and upper case letters in bold face
($\ve{a}$, $\m{A}$) to express vectors and matrices, respectively. A tilde
($\vef{a}$, $\mf{A}$) shall indicate frequency domain variables. We address
with $a[k]$ element $k$ of a vector $\ve{a}$, $[\m{A}]_{k,l}$ refers to
the element of the $k$th row and $l$th column, $[\m{A}]_{k,*}$ corresponds to all
elements of row $k$, and $[\m{A}]_{*,l}$  all
elements of column $l$. We use $(\cdot)^T$ for the transpose,
$(\cdot)^H$ for the conjugate transpose or
Hermitian, $\E{\cdot}$ for expectation, $\diag{\m{A}}$ for the main
diagonal elements of $\m{A}$, $\tr{\m{A}}$ to express the trace
operator, and $(\cdot)^\dagger$ to denote a Moore-Penrose
Pseudo-Inverse. Identity and zero matrices are denoted as $\m{I}$ and 
$\m{0}$, respectively. A vector $\ve{a}\sim\mathcal{CN}\left(\upmuvec,\m{C}\right)$
denotes a circularly symmetric complex Gaussian noise vector with mean
$\upmuvec$ and covariance matrix $\m{C}$. We denote an estimation of $\ve{a}$
as $\hat{\ve{a}}$. We indicate the \underline{m}otivation behind a
specific subscript/superscript of $\ve{a}_m$ by underlining a letter
accordingly. Representation in equivalent complex baseband applies to all
signals and systems.



\section{UW-OFDM Signaling Model}\label{sec:review}
\begin{figure}[htb]
\centering
\begin{tikzpicture}
	\clip (-25, -7) rectangle (50, 17);
	\draw (0, 0) rectangle +(30, 8);
	\draw[pattern=horizontal lines light gray] (30, 0) rectangle +(10, 8);
	\node at (35, 4){UW};
	\draw[|<->|] (30, 12) -- node[above] {$N_g$} +(10, 0);
	\draw[|<->|] (0, 10) -- node[above] {$N$} +(40, 0);
	\draw (-40, 0) rectangle +(30, 8);
	\draw[pattern=horizontal lines light gray] (-10, 0) rectangle +(10, 8);
	\node at (-5, 4){UW};
	\draw (40, 0) rectangle +(30, 8);
	\draw[|<->|] (0, -2) -- node[below] {UW-OFDM symbol} +(40, 0);
\end{tikzpicture}
\caption{Structure of an UW-OFDM time domain symbol.}
\label{fig:uwofdm_symbol}
\end{figure}
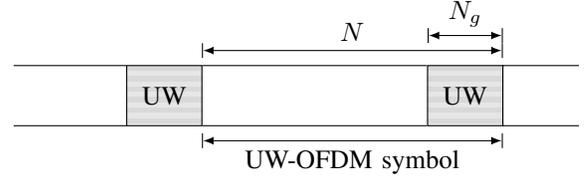

In this section, we briefly review the UW-OFM signaling model. For details, the
interested reader is referred to \cite{Hofbauer20}.
Let $\ve{x}_u\in\mathbb{C}^{N_u \times 1}$ denote the
\underline{U}W, which we use to fill the guard interval of length $N_g=N_u$
and that is part of each OFDM time domain symbol of size $N$ (see Fig.~\ref{fig:uwofdm_symbol}).
We therefore introduce $\xt=\begin{bmatrix}\ve{x}_{\textpl}^T & \ve{x}_u^T \end{bmatrix}^T$, $\xt\in
\mathbb{C}^{N \times 1}$, whereas $\ve{x}_{\textpl}\in\mathbb{C}^{(N-N_u) \times 1}$ carries the
\underline{p}ay\underline{l}oad. Following transmission energy arguments 
in \cite{Huemer14}, we obtain $\xt$ by generating $\ve{x}
= \begin{bmatrix}\ve{x}_{\textpl}^T & \ve{0}^T\end{bmatrix}^T$ first and then
  add the UW with $\xt =
  \ve{x} + \begin{bmatrix}\ve{0}^T & \ve{x}_u^T\end{bmatrix}^T$.
As in conventional OFDM, unused $N_z$ \underline{z}ero subcarriers for spectral
shaping reasons, $N_p$ pilot symbols $\ve{p}\in \mathbb{C}^{N_p\times 1}$ for
estimation purposes, and $N_d$ data symbols
$\ve{d}\in\mathcal{A}^{N_d \times 1}$ drawn from an alphabet
$\mathcal{A}$ form an OFDM frequency domain symbol $\vef{x}\in\mathbb{C}^{N \times 1}$. In order to account for the
zero-word of the UW-OFDM time domain symbol $\ve{x}$ and
thus fulfill the system of
equations
$\m{F}_N^{-1}\vef{x}=\begin{bmatrix}\ve{x}_{\textpl}^T & \ve{0}^T\end{bmatrix}^T$,
whereas $\m{F}_{N}^{-1}$ is an $N$-point inverse DFT with
$\m{F}_{N}^{-1} = \frac{1}{N}\m{F}_{N}^H$ and $\mel{\m{F}_N}_{k,l} =
\e^{-j\frac{2\pi}{N}kl}$, we have to reduce $N_d$ by at least $N_u$, and instead
introduce some form of redundancy.
We therefore define a 
\underline{d}ata generator matrix $\m{G}_d\in\mathbb{C}^{(N_d+N_r+N_p)\times N_d}$ and
and a \underline{p}ilot generator matrix
$\m{G}_p\in\mathbb{C}^{(N_d+N_r+N_p)\times N_p}$ with $N_r=N_u$ and $N =
N_d+N_r+N_p+N_z$, yielding
\begin{align}
\ve{x}=\m{F}_N^{-1}\left(\m{B}\m{G}_d\ve{d} +
\m{B}\m{G}_p\ve{p}\right)  = \begin{bmatrix} \ve{x}_{\textpl}
  \\ \ve{0} \end{bmatrix},\label{equ:pil003} 
\end{align}
while $\m{B}\in\{0,1\}^{N \times (N-N_z)}$ accounts for the zero subcarrier insertion.

\begin{figure}[!t]
\centering
\subfloat[$|\m{G}'_d|$]{\scalebox{0.3}{
%
\begin{tikzpicture}

\begin{axis}[%
width=4.047in,
height=3.566in,
at={(0.679in,0.481in)},
scale only axis,
point meta min=0,
point meta max=0.841553421417078,
axis on top,
xmin=0.5,
xmax=32.5,
y dir=reverse,
ymin=0.5,
ymax=52.5,
axis background/.style={fill=white},
legend style={legend cell align=left, align=left, draw=white!15!black},
colormap={mymap}{[1pt] rgb(0pt)=(1,1,1); rgb(64pt)=(0,0,0)},
colorbar
]
\addplot [forget plot] graphics [xmin=0.5, xmax=32.5, ymin=0.5, ymax=52.5] {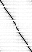};
\end{axis}
\end{tikzpicture}%
}
\label{fig:Gd1_abs}}
\hfill
  \subfloat[$|\m{G}''_d|$]{\scalebox{0.3}{
%
\begin{tikzpicture}

\begin{axis}[%
width=4.047in,
height=3.566in,
at={(0.679in,0.481in)},
scale only axis,
point meta min=0,
point meta max=0.404388521268797,
axis on top,
xmin=0.5,
xmax=32.5,
y dir=reverse,
ymin=0.5,
ymax=52.5,
axis background/.style={fill=white},
legend style={legend cell align=left, align=left, draw=white!15!black},
colormap={mymap}{[1pt] rgb(0pt)=(1,1,1); rgb(64pt)=(0,0,0)},
colorbar
]
\addplot [forget plot] graphics [xmin=0.5, xmax=32.5, ymin=0.5, ymax=52.5] {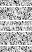};
\end{axis}
\end{tikzpicture}%
}  
\label{fig:Gd2_abs}}
\caption{Exemplary data generator matrices $\m{G}'_d$  and $\m{G}''_d$.}
\label{fig:Gd_abs}
\end{figure}

Matrix $\m{G}_d$ in \eqref{equ:pil003} maps the data symbols $\ve{d}$ on the
UW-OFDM symbol, while at the same time accounting for the fulfillment of the zero-word
constraint. There are several degrees of freedom in designing $\m{G}_d$, yielding realizations with different properties (see details in
\cite{Huemer12_1,Hofbauer20}). We thus show two exemplary realizations of $\m{G}_d$
in  Fig.~\ref{fig:Gd_abs} to indicate the comprehensive set of generator matrices, and
we will use them as a basis for discussions in the subsequent sections. Matrix
$\m{G}'_d$ in Fig.~\ref{fig:Gd1_abs} has (energy relevant) entries primarily on
the main diagonal. Consequently, a single data symbol is mainly mapped onto a
single subcarrier, however, some parts  are also spread on neighboring
subcarriers, which in turn will generate the zero-word. The design and also resulting behavior
of $\m{G}'_d$ is similar to a conventional OFDM system, which would in the
given signal framework represented by a matrix
solely with main diagonal entries. In contrast, $\m{G}''_d$ shown in Fig.~\ref{fig:Gd2_abs}
spreads each data symbol almost uniformly over all subcarriers and thus behaves similar to a
single-carrier (SC) based system.

\begin{figure}[!t]
  \centering
  \scalebox{0.4}{
%
\begin{tikzpicture}

\begin{axis}[%
width=2.195in,
height=3.566in,
at={(0.368in,0.481in)},
scale only axis,
point meta min=0,
point meta max=1,
axis on top,
xmin=0.5,
xmax=4.5,
xtick={1,2,3,4},
xticklabels={{0},{1},{2},{3}},
xlabel style={font=\color{white!15!black}},
xlabel={$\left|\bf{G}_p\right|$},
y dir=reverse,
ymin=0.5,
ymax=52.5,
axis background/.style={fill=white},
legend style={legend cell align=left, align=left, draw=white!15!black},
colormap={mymap}{[1pt] rgb(0pt)=(1,1,1); rgb(64pt)=(0,0,0)},
colorbar
]
\addplot [forget plot] graphics [xmin=0.5, xmax=4.5, ymin=0.5, ymax=52.5] {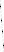};
\end{axis}
\end{tikzpicture}%
}
\caption{Exemplary pilot generator matrix $\m{G}_p$.}
\label{fig:Gp_abs}
\end{figure}

Matrix $\m{G}_p$ in \eqref{equ:pil003} maps the pilot
symbols $\ve{p}$ on the UW-OFDM symbol, again accounting for the generation of the zero-word (see \cite{Hofbauer20} for details on the design). Fig.~\ref{fig:Gp_abs}
shows the resulting $\m{G}_p$ for the setup given in Tab.~\ref{tab:setups} in Sec.~\ref{sec:setup}.

With \eqref{equ:pil003}, the
$l$th frequency domain UW-OFDM transmit signal follows as
\begin{equation}
  \Xfl = \m{F}_N\xtl = \m{B}\m{G}_d\datal + \m{B}\m{G}_p\ve{p} + \vef{x}_u,
\end{equation}
whereas $\vef{x}_u=\m{F}_N\begin{bmatrix}\ve{0}^T &
\ve{x}_u^T\end{bmatrix}^T$ denotes the UW in the frequency domain.

The OFDM frequency domain symbol at the
\underline{r}eceiver $\Yfrxl$ can be modelled as\footnote{In this work, the notation $'$
  serves either as a naming convention for different generator
  matrix instances or to differentiate between matrices incorporating all
  or only the subset of non-zero subcarriers (e.g., $\mf{H}'$ versus
  $\mf{H}'$).}
\begin{align}
\Yfrxl &= \lambdaNfl\mf{H}'\Xfl + \noiseffull,\label{equ:cfo011}
\end{align}
with $\mf{H}'\in\mathbb{C}^{N\times N}$ corresponding to the diagonal channel frequency
response matrix, $\noiseffull=\sim\mathcal{CN}(\ve{0},\varnoisef\m{I})$, $\noiseffull\in\mathbb{C}^{N \times 1}$, and
the frequency domain representation of the CFO effects given as \cite{Hofbauer20}
\begin{align}
  \lambdaNfl = \e^{j\varphil}\mf{\Lambda}'_\text{stat},\, \lambdaNfl\in\mathbb{C}^{N\times N}\label{equ:cfo101}
\end{align}
and
\begin{align}
 \varphil &= \frac{2\pi\epsilon}{N}\left(Nl+\frac{N-1}{2}+N_x\right)\label{equ:cfo103},\\ \mel{\mf{\Lambda}'_\text{stat}}_{k,m}&=\frac{\sin\left(\pi(m+\epsilon-k)\right)}{N\sin\left(\frac{\pi(m+\epsilon-k)}{N}\right)}\e^{j\frac{\pi(m-k)(N-1)}{N}}.\label{equ:cfo031}
\end{align}
Here, $\epsilon=\tfrac{f_\cfo}{\Delta_f}$ denotes a CFO $f_\cfo$
normalized to the subcarrier spacing $\Delta_f$, and $\varphil$ carries the phase
offset accumulated by previous OFDM symbols up to and including symbol $l$ and an arbitrary offset $N_x\in\mathbb{Z}$.
In summary, a frequency domain symbol suffers from a phase
offset $\varphil$ also referred to as CPE, an attenuation determined by the entries on the main diagonal of $\mf{\Lambda}'_\text{stat}\in\mathbb{C}^{N\times N}$,
and a deviation as a result of intercarrier interference (ICI) represented by
the off-diagonal entries of $\mf{\Lambda}'_\text{stat}$. The notation \underline{stat}ic stems from the independence of $l$.

\section{CFO Estimation and Compensation in UW-OFDM}
\label{sec:impact_cfo_mse}
In the following, we will evaluate the impact of CFO
effects on the UW-OFDM performance. For that we will apply different CFO
compensation methods and evaluate the performance by conducting an MSE
analysis after data estimation. Please
note that CFO compensation by means of CPE correction presented in
Sec.~\ref{sec:cpe_correction} has already been part in the conference version
of this work \cite{Hofbauer20_2}. However, we included it in a compressed form as a frame of
reference for the advanced compensation methods in
Sec.~\ref{sec:advanced_cfo_comp}, as well as for the derivations of the error
model in Sec.~\ref{sec:error_model}.


We start with a removal of the zero subcarriers and extract the
\underline{p}ay\underline{l}oad carrying subcarriers, which can be well
approximated\footnote{We neglect the potential leakage of $\vef{x}_u$ parts from the zero to
  the non-zero subcarriers due to ICI.} by \cite{Hofbauer20} 
\begin{align}
  \Yfdownl &= \m{B}^T\Yfrxl\label{equ:cfo_rx020}\\
&\approx \lambdafl\mf{H}\m{G}_d\datal +
\lambdafl\mf{H}\m{G}_p\ve{p} + \lambdafl\mf{H}\m{B}^T\vef{x}_u + \noisef,\nonumber
\end{align}
where $\noisef=\m{B}^T\noiseffull\sim\mathcal{CN}(\ve{0},\varnoisef\m{I})$,
$\noisef\in\mathbb{C}^{(N-N_z) \times 1}$, $\mf{H}=\m{B}^T\mf{H}'\m{B}$,
$\mf{H}\in\mathbb{C}^{(N-N_z)\times (N-N_z)}$,
$\lambdafl=\m{B}^T\lambdaNfl\m{B}$, and $\lambdafl\in\mathbb{C}^{(N-N_z)\times
  (N-N_z)}$. We note that $\mf{H}$ and $\lambdafl$ coincide with 
$\mf{H}'$ and $\lambdaNfl$, respectively, except for excluded rows and columns at the
positions of zero subcarriers. In the following, this signaling model serves as a
starting point for the evaluation of different CFO compensation methods.

\subsection{CPE correction}\label{sec:cpe_correction}

A common means to combat CFO impairments is CPE correction by derotating with
an estimate $\varphilest$. Together with the subtraction of the offset
\begin{align}
\ve{x}_\text{off} = \mf{H}\m{G}_p\ve{p} + \mf{H}\m{B}^T\vef{x}_u
\end{align}
from \eqref{equ:cfo_rx020} we yield
\begin{align}
  \begin{split}
  \Yfl &= \e^{-j\varphilest}\Yfdownl - \ve{x}_\text{off}\\
&= \e^{-j\varphilest}\lambdafl\mf{H}\m{G}_d\datal +\left(\e^{-j\varphilest}\lambdafl-\m{I}\right)\mf{H}\m{G}_p\ve{p}\\
&\quad+
  \left(\e^{-j\varphilest}\lambdafl-\m{I}\right)\mf{H}\m{B}^T\vef{x}_u
  + \e^{-j\varphilest}\noisef
\end{split}\label{equ:cfo_rx023}\\
  &= \e^{-j\varphilest}\lambdafl\mf{H}\m{G}_d\datal + \errl +\e^{-j\varphilest}\noisef,\label{equ:cfo_rx024a}
\end{align}
with $\errl$ denoting the residual UW and pilot offset. Assuming that CPE correction already
compensates the CFO effects sufficiently, i.e.,
$\e^{-j\varphilest}\lambdafl\approx\m{I}$, we have $\errl\approx\ve{0}$ and
obtain a simplified linear system model
\begin{align}
  \Yfl \approx \mf{H}\m{G}_d\datal +\e^{-j\varphilest}\noisef. \label{equ:rx_simple_system_model}
\end{align}
For this system model and with $\E{\datal{\datal}^H}=\sigma_d^2\m{I}$, a linear
minimum mean square error (LMMSE) estimator \cite{Kay93} given as
\begin{align}
\m{E}_\lmmse=(\m{G}_d^H\mf{H}^H \mf{H}\m{G}_d + \tfrac{N
  \varnoisef}{\sigma_d^2}\m{I})^{-1} \m{G}_d^H\mf{H}^H\label{equ:lmmse_estimator}
\end{align}
is a good choice to estimate the data symbols as
\begin{equation}
\dataestl=\m{E}_\lmmse\Yfl.\label{equ:cfo_rx150}
\end{equation}
The $k$th element of $\dataestl$ follows as 
\begin{align}
\kElemdataestl &=
\e^{-j\varphilest}\e^{j\varphil}\ve{e}_k^T\mf{\Lambda}_\text{stat}\mf{H}\ve{g}_k\kElemdatal+\Delta_k+\e^{-j\varphilest}\ve{e}_k^T\noisef,\label{equ:cfo_rx031a}
\end{align}
with $\ve{e}_k^T=\mel{\m{E}_\text{LMMSE}}_{k,*}$,
$\ve{g}_k=\mel{\m{G}_d}_{*,k}$,
$\mf{\Lambda}_\text{stat}=\m{B}^T\mf{\Lambda}'_\text{stat}\m{B}$, $\mf{\Lambda}_\text{stat}\in\mathbb{C}^{(N-N_z)\times (N-N_z)}$, and
\begin{align}
\Delta_k=\e^{-j\varphilest}\e^{j\varphil}\ve{e}_k^T\mf{\Lambda}_\text{stat}\mf{H}\sum_{m=0,m\neq
  k}^{N_d-1}\ve{g}_m\mElemdatal.\label{equ:cfo100}
\end{align}
The estimate $\kElemdataestl$ in \eqref{equ:cfo_rx031a} consists of a corrupted
version of $\kElemdatal$, an impairment by the other data
symbols $\Delta_k$ due to ICI, and an additive noise
term. We keep the latter for model completeness to be available for further
utilization in subsequent sections. However, in order to reveal the CFO effects, we set
$\noisef=\ve{0}$ for the following considerations. Before we conduct MSE
analyses on $\dataestl$, let us first determine the impact of $\varphilest$ on
$\dataestl$ by examining two CPE estimator variants.

\subsubsection{Perfect CPE estimation}
We start with a perfect CPE estimator given as
$\varphilest=\varphil$. Next, we introduce a magnitude phase model
$a_{\datasscript,k}\e^{j\varphi_{\datasscript,k}} =
\ve{e}_k^T\mf{\Lambda}_\text{stat}\mf{H}\ve{g}_k$ in \eqref{equ:cfo_rx031a} to
represent the impairment on the $k$th \underline{d}ata symbol as
\begin{align}
  \kElemdataestlperfect = a_{\datasscript,k}\e^{j\varphi_{\datasscript,k}}
  \kElemdatal+\Delta_k\approx \e^{j\varphi_{\datasscript,k}}
  \kElemdatal+\Delta_k.\label{equ:cfo_150}
\end{align}
The approximation $a_{\datasscript,k}\approx 1$ relies on an estimator
cancelling out the impact of the channel and generator matrix such that $\ve{e}_k^T\mf{H}\ve{g}_k\approx 1$, and the approximation
$\left|\mf{\Lambda}_\text{stat}\right|\approx\m{I}$ \cite{Hofbauer16}.
\begin{figure}[!t]
\centering
\scalebox{0.66}{\input{./fig/const_diag_perf_phase_est.tex}}
\caption{Constellation diagram of $\dataestl$ for $\varphilest=\varphil$, QPSK,
$\mf{H}=\m{I}$, $\vef{x}_u=\ve{0}$, $\errl=\ve{0}$, $\epsilon=0.1$, $l=0,\ldots, L-1$, and $\varnoisef=0$.}
\label{fig:const_diag_perf_vs_imperf_phase_est}
\end{figure}
The estimate $\kElemdataestlperfect$ in \eqref{equ:cfo_150} deviates from the
transmitted symbols $\kElemdatal$ by $\varphi_{\datasscript,k}$ and
$\Delta_{k}$, causing a phase rotation and an expansion of single data points to
clouds, respectively. Both, UW-OFDM and CP-OFDM, suffer from $\Delta_{k}$
(although with different power levels), but the existence of $\varphi_{\datasscript,k}$ is limited to UW-OFDM only,
cf.~Fig.~\ref{fig:const_diag_perf_vs_imperf_phase_est}. It orginates from the
combination of $\m{G}_d$ distributing a data symbol across neighboring subcarriers and ICI leaking back parts
of it, thus resulting in a self-interference of $\kElemdatal$.

\subsubsection{Non-perfect pilot tone based CPE estimation}
Following \cite{Hofbauer20}, an
estimate $\varphilest$ based on the pilot tones $\ve{p}$ is given as
\begin{equation}
\varphilest = \varphil + \varphi_{\pil} + \delta_l,\label{equ:cfo_varphilest}
\end{equation}
which deviates from $\varphil$ by an additive random deviation $\delta_l$ and
a phase offset $\varphi_{\pil}$ caused by self interference of the \underline{p}ilot
symbols. The latter originates from the same underlying effect as
$\varphi_{\datasscript,k}$. Incorporating
\eqref{equ:cfo_varphilest} into \eqref{equ:cfo_rx031a} yields
\begin{align}
\kElemdataestlnonperfect \approx
\e^{-j(\varphil +\varphi_\pil+ \delta_l)}\e^{j\varphil}\e^{j\varphi_{\datasscript,k}}\kElemdatal
+ \Delta_k
\approx \e^{j\varphi_\text{off}}\kElemdatal,\label{equ:cfo_rx037}
\end{align}
with a remaining and approximated \underline{off}set
$\varphi_{\text{off}} =\frac{1}{N_d}\sum_{k=0}^{N_d-1}\varphi_{\datasscript,k}
-\varphi_\pil=\varphi_{\datasscript} -\varphi_\pil$. For this approximation, we
neglect the minor impact of $\delta_l$ and use an averaged offset
$\varphi_\text{off}$ with $\varphi_{\datasscript}$ for all subcarriers, as
there is almost no
performance difference compared to subcarrier individual offsets with $\varphi_{\datasscript,k}$ \cite{Hofbauer16}.
We refine the
estimate in \eqref{equ:cfo_rx150} by incorporating the effect of $\varphi_\text{off}$. Additionally, we refrain from the approximation
$\e^{-j\varphilest}\lambdafl\approx\m{I}$, which is the underlying basis for the
simplified model of $\Yfl$ in \eqref{equ:rx_simple_system_model}, yielding now
\begin{align}
  \dataestl&=\e^{-j\hat{\varphi}_\text{off}}\m{E}_\text{LMMSE}\Yfl\label{equ:cfo_rx039}\\
  &= \e^{-j\hat{\varphi}_\text{off}}\e^{-j\varphilest}\m{E}_\text{LMMSE}\lambdafl\mf{H}\m{G}_d\datal + \e^{-j\hat{\varphi}_\text{off}}\m{E}_\text{LMMSE}\errl\nonumber\\ &\quad+\e^{-j\hat{\varphi}_\text{off}}\e^{-j\varphilest}\m{E}_\text{LMMSE}\noisef\label{equ:cfo_rx040}.
\end{align}
We obtain an estimate $\est{\varphi}_\text{off}$ according to
\begin{align}
\est{\varphi}_\text{off}&=\est{\varphi}_{\datasscript}-\est{\varphi}_\pil=\left(m_{\datasscript}-m_\pil\right)\est{\epsilon},\label{equ:est_varphi_off}
\end{align}
with $m_{\datasscript},m_\pil\in\mathbb{R}$ derived from numerical
evaluations that depend on the UW-OFDM setup and $\mf{H}$, and an estimate
$\est{\epsilon}$ based on \eqref{equ:cfo103} with the
details in \cite{Hofbauer20}.


For the subsequent evaluations, we use the same systems and assume the same multipath
channels as propagation
environment as for the BER
simulations later on, with the setups detailed in Sec.~\ref{sec:setup}. We
utilize a Barker code as $\ve{x}_u$ \cite{Golomb1965} for the non-zero
case, and scale the samples to yield equal average power compared to the rest
of the symbol. We estimate $\varphilest$ based on pilot tones
\cite{Hofbauer20} with the error given in \eqref{equ:cfo_varphilest}, and
$\est{\epsilon}$ once per packet \cite{Hofbauer20} based on \eqref{equ:cfo103}.

Fig.~\ref{fig:mse_cfo_1} shows the Bayesian MSE (BMSE) per data symbol $\theta_d
=\frac{1}{L}\sum_{l=0}^{L-1}\frac{1}{N_d}\E{\left\|\dataestl-\datal\right\|_2^2}$
as a function of the CFO $\epsilon$.
\begin{figure}[!t]
\centering
\scalebox{0.66}{
%
\definecolor{mycolor1}{rgb}{0.43529,0.72941,0.23922}%
\definecolor{mycolor2}{rgb}{1.00000,0.50196,0.00000}%
\definecolor{mycolor3}{rgb}{0.12941,0.44314,0.75686}%
\begin{tikzpicture}

\begin{axis}[%
width=4.521in,
height=3.566in,
at={(0.758in,0.481in)},
scale only axis,
xmin=0,
xmax=0.1,
xtick={0,0.02,0.04,0.06,0.08,0.1},
xticklabels={{$0$},{$0.02$},{$0.04$},{$0.06$},{$0.08$},{$0.1$}},
xlabel style={font=\color{white!15!black}},
xlabel={carrier frequency offset $\epsilon$},
ymin=0,
ymax=0.1,
ytick={0,0.01,0.02,0.03,0.04,0.05,0.06,0.07,0.08,0.09,0.1},
yticklabels={{$0$},{$0.01$},{$0.02$},{$0.03$},{$0.04$},{$0.05$},{$0.06$},{$0.07$},{$0.08$},{$0.09$},{$0.1$}},
ylabel style={font=\color{white!15!black}},
ylabel={BMSE $\theta_d$},
axis background/.style={fill=white},
xmajorgrids,
ymajorgrids,
legend style={at={(0.03,0.97)}, anchor=north west, legend cell align=left, align=left, draw=white!15!black}
]
\addplot [color=black, line width=1.0pt, mark=square, mark options={solid, black}]
  table[row sep=crcr]{%
0	1.05560613680834e-30\\
0.01	0.000958085084740888\\
0.02	0.00383040202381421\\
0.03	0.0085782507820361\\
0.04	0.0152177382496837\\
0.05	0.0238534777531106\\
0.06	0.0341065027017116\\
0.07	0.0465999580478759\\
0.08	0.0603337416527625\\
0.09	0.0760744463577694\\
0.1	0.0941870341425964\\
};
\addlegendentry{$\m{G}_{d,\cpofdm}$, compens. by $\varphilest$}

\addplot [color=mycolor1, dashdotted, line width=1.0pt, mark=triangle, mark options={solid, rotate=270, mycolor1}]
  table[row sep=crcr]{%
0	1.51500245889634e-30\\
0.01	0.000305345661018082\\
0.02	0.00122103583386898\\
0.03	0.0027500664961881\\
0.04	0.00488818259741404\\
0.05	0.00763460238352742\\
0.06	0.011002830522827\\
0.07	0.0149597533280509\\
0.08	0.0195451882838427\\
0.09	0.0247292167803165\\
0.1	0.0305234078557593\\
};
\addlegendentry{$\m{G}'_d$, $\ve{x}_u\neq\ve{0}$, compens. by $\varphilest$}

\addplot [color=mycolor1, line width=1.0pt, mark=triangle*, mark options={solid, rotate=270, fill=mycolor1, mycolor1}]
  table[row sep=crcr]{%
0	1.51512071869865e-30\\
0.01	0.000239738330925971\\
0.02	0.000959014581106974\\
0.03	0.00216002657783102\\
0.04	0.00384251778086366\\
0.05	0.00600854960466046\\
0.06	0.00865823992641523\\
0.07	0.0117860674064888\\
0.08	0.0153996830702456\\
0.09	0.0195100328030495\\
0.1	0.0240924226093609\\
};
\addlegendentry{$\m{G}'_d$, $\ve{x}_u\neq\ve{0}$, compens. by $\varphilest$, $\est{\varphi}_\text{off}$}

\addplot [color=mycolor2, dashdotted, line width=1.0pt, mark=square, mark options={solid, mycolor2}]
  table[row sep=crcr]{%
0	1.30086462269532e-29\\
0.01	0.000202892054194911\\
0.02	0.000812140111509199\\
0.03	0.0018268062275786\\
0.04	0.00325092164998623\\
0.05	0.00508115565705213\\
0.06	0.00731876178658663\\
0.07	0.00996263796149711\\
0.08	0.0130182951735933\\
0.09	0.0164767074257681\\
0.1	0.0203457699023341\\
};
\addlegendentry{$\m{G}'_d$, $\ve{x}_u=\ve{0}$, compens. by $\varphilest$}

\addplot [color=mycolor2, line width=1.0pt, mark=square*, mark options={solid, fill=mycolor2, mycolor2}]
  table[row sep=crcr]{%
0	1.30086462269532e-29\\
0.01	0.00019555895732814\\
0.02	0.000782787982582437\\
0.03	0.00176095037527299\\
0.04	0.00313407192337092\\
0.05	0.00489758020324535\\
0.06	0.00705497890183359\\
0.07	0.00960531809243242\\
0.08	0.0125495122018382\\
0.09	0.0158843637414357\\
0.1	0.0196165654267624\\
};
\addlegendentry{$\m{G}'_d$, $\ve{x}_u=\ve{0}$, compens. by $\varphilest$, $\varphi_\text{off}$}

\addplot [color=mycolor3, dashdotted, line width=1.0pt, mark=o, mark options={solid, mycolor3}]
  table[row sep=crcr]{%
0	3.26240562512445e-29\\
0.01	0.00020298899219569\\
0.02	0.000812780072504064\\
0.03	0.00182888343982264\\
0.04	0.00325100802319403\\
0.05	0.00508569514754654\\
0.06	0.00731979156667763\\
0.07	0.00997005158981123\\
0.08	0.0130288528243068\\
0.09	0.0164970438066337\\
0.1	0.0203681477596526\\
};
\addlegendentry{$\m{G}''_d$, $\ve{x}_u=\ve{0}$, compens. by $\varphilest$}

\addplot [color=mycolor3, line width=1.0pt, mark=*, mark options={solid, fill=mycolor3, mycolor3}]
  table[row sep=crcr]{%
0	3.26240562512445e-29\\
0.01	0.000197308160400436\\
0.02	0.000790040694182567\\
0.03	0.00177750574065286\\
0.04	0.00315993273201782\\
0.05	0.00494358470280423\\
0.06	0.00711587185769639\\
0.07	0.00969073732849286\\
0.08	0.0126638475763461\\
0.09	0.0160375398895491\\
0.1	0.0198001961163145\\
};
\addlegendentry{$\m{G}''_d$, $\ve{x}_u=\ve{0}$, compens. by $\varphilest$, $\varphi_\text{off}$}

\end{axis}
\end{tikzpicture}%
}
\caption{BMSE $\theta_d$ of $\dataestl$ from
  \eqref{equ:cfo_rx150} and \eqref{equ:cfo_rx039} for UW-OFDM and
  CP-OFDM ($\m{G}_{d,\cp}$) with $l=0,\ldots L-1$, $\sigma_d^2=1$, and
  $\varnoisef=0$.}
\label{fig:mse_cfo_1}
\end{figure}
UW-OFDM outperforms CP-OFDM ($\m{G}_{d,\cpofdm}$) for all investigated setups,
which is due to the reduced intercarrier/interdata effects compared to CP-OFDM,
and follows from a property inherited from the special UW-OFDM generator matrix structures \cite{Hofbauer20}. 

The overlapping and two most lower curves in Fig.~\ref{fig:mse_cfo_1} (solid
blue line for $\m{G}'_d$ and solid orange line for $\m{G}''_d$) represent
bounds for an optimal performance in this scenario (i.e., zero UW, $\errl=\ve{0}$ and $\hat{\varphi}_\text{off}=\varphi_\text{off}$). Comparing with the case without
offset compensation as depicted by the
(overlapping) dashed lines slightly above, we conclude that the gain of the latter is rather
limited in the zero UW case.

Obviously, the performance gain due to phase offset compensation is more dominant in the
non-zero UW case. It increases with the CFO value and is almost the same
regardless of compensating by $\varphi_\text{off}$ or only by an estimate
$\hat{\varphi}_\text{off}$ (thus only the latter is shown). For the non-zero UW
case, results are limited to $\m{G}'_d$, as the zero UW
scenario already confirmed a similar CFO robustness of $\m{G}'_d$ and $\m{G}''_d$.

The most lower curve in Fig.~\ref{fig:mse_cfo_1} confirms that
even in the best case of CPE correction,
there is still room for improvement due to an incomplete CFO cancelation. Closing the remaining performance gap will thus be
tackled in the subsequent section.


\subsection{Advanced CFO Compensation Techniques}
\label{sec:advanced_cfo_comp}
Next, we aim at decreasing the MSE that remains after applying the
methods presented in the preceding section even further, relying on the methods
proposed in \cite{Hofbauer16}. With \eqref{equ:cfo_rx024a}, $\ve{x}_u=\ve{0}$
and with $\errl=\ve{0}$, the system model can be written as
\begin{equation}
\Yfl = \e^{-j\varphilest}\lambdafl\mf{H}\m{G}_d\datal+ \e^{-j\varphilest}\noisef.\label{equ:cfo_comp001}
\end{equation}
With an estimate $\varphilest$ according to
\eqref{equ:cfo_varphilest}, we obtain
\begin{align}
\Yfl &=
\e^{-j(\varphil+\varphi_\pil+\delta_l)}\lambdafl\mf{H}\m{G}_d\datal+
\e^{-j(\varphil+\varphi_\pil+\delta_l)}\noisef\label{equ:cfo_comp002}\\
&=\e^{-j(\varphi_\pil+\delta_l)}\mf{\Lambda}_\text{stat}\mf{H}\m{G}_d\datal+ \e^{-j(\varphil+\varphi_\pil+\delta_l)}\noisef.\label{equ:cfo_comp003}
\end{align}
Aside from $\delta_l$ (which has only a minor impact \cite{Hofbauer20}), all OFDM symbols experience the same remaining CFO impairments, which can be
fully compensated by multiyplying with
$\e^{j\varphi_\pil}\mf{\Lambda}_\text{stat}^{-1}$. As $\varphi_\pil$
and $\mf{\Lambda}_\text{stat}$ depend on $\epsilon$, only
estimates of them are available. Applying
$\e^{j\est{\varphi}_\pil}\est{\mf{\Lambda}}_\text{stat}^{-1}$ on $\Yfl$
followed by $\m{E}_\lmmse$ from
\eqref{equ:lmmse_estimator}, a data estimate is obtained as
\begin{align}
\dataestl &= \m{E}_\lmmse\e^{j\hat{\varphi}_\pil}\hat{\mf{\Lambda}}_\text{stat}^{-1}\Yfl\\
\begin{split} &=\e^{j(\hat{\varphi}_\pil-\varphi_\pil-\delta_l)}\m{E}_\lmmse\hat{\mf{\Lambda}}^{-1}_\text{stat}\mf{\Lambda}_\text{stat}\mf{H}\m{G}_d\datal\\
  &\quad+ \e^{j(\hat{\varphi}_\pil-\varphi_\pil-\delta_l-\varphil)}\m{E}_\lmmse\hat{\mf{\Lambda}}^{-1}_\text{stat}\noisef.\label{equ:cfo_comp005a}
\end{split}
\end{align}
As a matrix inversion denotes an
additional, non-negligible computational overhead, we draw on the
similarities of $\mf{\Lambda}_\text{stat}$ with a unitary matrix,
cf.\,\eqref{equ:cfo031}, and approximate
 \begin{equation}
\mf{\Lambda}_\text{stat}^{-1} \approx \mf{\Lambda}_\text{stat}^{H},\label{equ:cfo_comp004}
 \end{equation}
 yielding
\begin{align}
  \begin{split}
    \dataestl &=\e^{j(\hat{\varphi}_\pil-\varphi_\pil-\delta_l)}\m{E}_\lmmse\hat{\mf{\Lambda}}^H_\text{stat}\mf{\Lambda}_\text{stat}\mf{H}\m{G}_d\datal\\
  &\quad+
    \e^{j(\hat{\varphi}_\pil-\varphi_\pil-\delta_l-\varphil)}\m{E}_\lmmse\hat{\mf{\Lambda}}^H_\text{stat}\noisef.\label{equ:cfo_comp005}\end{split}
\end{align}

Fig.~\ref{fig:mse_cfo_3} confirms for $\m{G}'_d$ a significant gain by applying
the proposed advanced CFO compensation methods. Please note that we observed
very similar results for $\m{G}''_d$, but omit them here for the sake of clarity. The remaining gap to the performance bound with perfect
conditions, i.e., $\mf{\Lambda}_\text{stat}^{-1}$, $\varphilest$, and
$\varphi_\pil$ (note that $\varphi_\pil$ compensates for
$\varphilest\neq\varphil$ up to $\delta_l$), shows only a minor performance
degradation due to the applied approximations. Offset compensation becomes more
relevant compared to the results in Sec.~\ref{sec:cpe_correction}, as $\varphi_\pil$ and
$\varphi_\datasscript$ do not cancel each other as it is partially the case in
\eqref{equ:cfo_rx040}. In absolute terms, the gain due to additional offset compensation
is approximately doubled when comparing to simple CPE correction  by
$\varphilest$ (Fig.~\ref{fig:mse_cfo_1}). Also CP-OFDM benefits significantly
from the advanced methods and the gap to UW-OFDM is reduced
in comparison to Fig.~\ref{fig:mse_cfo_1}. However, UW-OFDM remains the better scheme.
\begin{figure}[!tbh]
\centering
\scalebox{0.66}{
%
\definecolor{mycolor1}{rgb}{1.00000,0.50196,0.00000}%
\begin{tikzpicture}

\begin{axis}[%
width=4.521in,
height=3.566in,
at={(0.758in,0.481in)},
scale only axis,
xmin=0,
xmax=0.1,
xtick={0,0.02,0.04,0.06,0.08,0.1},
xticklabels={{$0$},{$0.02$},{$0.04$},{$0.06$},{$0.08$},{$0.1$}},
xlabel style={font=\color{white!15!black}},
xlabel={carrier frequency offset $\epsilon$},
ymin=0,
ymax=0.02,
ylabel style={font=\color{white!15!black}},
ylabel={BMSE $\theta_d$},
axis background/.style={fill=white},
xmajorgrids,
ymajorgrids,
legend style={at={(0.03,0.97)}, anchor=north west, legend cell align=left, align=left, draw=white!15!black}
]
\addplot [color=black, line width=1.0pt, mark=o, mark options={solid, black}]
  table[row sep=crcr]{%
0	1.81236294692945e-28\\
0.01	5.98888152464799e-05\\
0.02	0.000239546156419066\\
0.03	0.000540609993057873\\
0.04	0.000961187420757413\\
0.05	0.00150364899665276\\
0.06	0.00217885254991559\\
0.07	0.00296841446694496\\
0.08	0.00388029414995139\\
0.09	0.00494410889872359\\
0.1	0.00615665845137805\\
};
\addlegendentry{$\m{G}_{d,\cpofdm}$, compens. by $\varphilest$, $\mf{\Lambda}_\text{stat}^{-1}$}

\addplot [color=black, dashdotted, line width=1.0pt, mark=+, mark options={solid, black}]
  table[row sep=crcr]{%
0	1.81417519541787e-28\\
0.01	7.96915573805877e-05\\
0.02	0.000344115168101603\\
0.03	0.000736792442486684\\
0.04	0.00129898628498825\\
0.05	0.00206021729842966\\
0.06	0.00299666888459669\\
0.07	0.00412948576357787\\
0.08	0.00539197885949013\\
0.09	0.00710184197929641\\
0.1	0.00877130045243127\\
};
\addlegendentry{$\m{G}_{d,\cpofdm}$, compens. by $\varphilest$, $\hat{\mf{\Lambda}}_\text{stat}^H$}

\addplot [color=mycolor1, line width=1.0pt, mark=square*, mark options={solid, fill=mycolor1, mycolor1}]
  table[row sep=crcr]{%
0	1.21982817145536e-29\\
0.01	0.000195575701982759\\
0.02	0.000782556020729239\\
0.03	0.00176110910360524\\
0.04	0.00313213967909365\\
0.05	0.00489581082872226\\
0.06	0.00705371096702134\\
0.07	0.00960055836231314\\
0.08	0.0125450328896605\\
0.09	0.0158886329183252\\
0.1	0.0196150589869491\\
};
\addlegendentry{$\m{G}'_d$, compens. by $\varphilest$, $\varphi_\text{off}$}

\addplot [color=mycolor1, line width=1.0pt, mark=triangle*, mark options={solid, rotate=270, fill=mycolor1, mycolor1}]
  table[row sep=crcr]{%
0	5.6306357946664e-29\\
0.01	8.50396202420823e-05\\
0.02	0.000340318473858684\\
0.03	0.000765954636444884\\
0.04	0.00136395834226531\\
0.05	0.00213761816408914\\
0.06	0.00308297990066082\\
0.07	0.00419663287050755\\
0.08	0.00548020953576582\\
0.09	0.00695973894851529\\
0.1	0.0086077625953895\\
};
\addlegendentry{$\m{G}'_d$, compens. by $\varphilest$, $\est{\mf{\Lambda}}_\text{stat}^H$}

\addplot [color=mycolor1, dashdotted, line width=1.0pt, mark=triangle, mark options={solid, rotate=270, mycolor1}]
  table[row sep=crcr]{%
0	5.60753874508172e-29\\
0.01	8.84640218069233e-05\\
0.02	0.000354464530223402\\
0.03	0.000796215497669804\\
0.04	0.00142285659859016\\
0.05	0.00222982774813841\\
0.06	0.00320198435812713\\
0.07	0.00435920543070996\\
0.08	0.00571549229709324\\
0.09	0.00724641803720874\\
0.1	0.00898797844434748\\
};
\addlegendentry{$\m{G}'_d$, compens. by $\varphilest$, $\mf{\Lambda}_\text{stat}^{-1}$}

\addplot [color=mycolor1, line width=1.0pt, mark=o, mark options={solid, mycolor1}]
  table[row sep=crcr]{%
0	5.6306357946664e-29\\
0.01	3.73397672886781e-05\\
0.02	0.000149675618104547\\
0.03	0.000337345044478201\\
0.04	0.000602678946690395\\
0.05	0.000944193314797526\\
0.06	0.00136697327919693\\
0.07	0.00186235388490702\\
0.08	0.00244378312892163\\
0.09	0.00311627651975937\\
0.1	0.00386121612806261\\
};
\addlegendentry{$\m{G}'_d$, compens. by $\varphilest$, $\varphi_\pil$, $\mf{\Lambda}_\text{stat}^{-1}$}

\addplot [color=mycolor1, dashdotted, line width=1.0pt, mark=+, mark options={solid, mycolor1}]
  table[row sep=crcr]{%
0	5.60753874508172e-29\\
0.01	4.18413855601685e-05\\
0.02	0.000167901373238533\\
0.03	0.000378122138731248\\
0.04	0.000675354663521603\\
0.05	0.00105908648040601\\
0.06	0.00153231801186506\\
0.07	0.00208676608460323\\
0.08	0.00274599055166381\\
0.09	0.0034939296293366\\
0.1	0.00434071559177724\\
};
\addlegendentry{$\m{G}'_d$, compens. by $\varphilest$, $\est{\varphi}_\pil$, $\est{\mf{\Lambda}}_\text{stat}^H$}

\end{axis}
\end{tikzpicture}%
}
\caption{BMSE $\theta_d$ of $\dataestl$ \eqref{equ:cfo_comp005} for advanced
  CFO compensation methods with $\ve{x}_u=\ve{0}$, $\errl=\ve{0}$, $l=0,\ldots
  L-1$, $\sigma_d^2=1$ and $\varnoisef=0$.}
\label{fig:mse_cfo_3}
\end{figure}

\section{Derivation of analytical error model for residual CFO impairments}\label{sec:error_model}

The compensation methods presented in Sec.~\ref{sec:impact_cfo_mse} are not
capable of perfectly canceling the CFO impairments, but will leave a certain residual error. In this section, we will derive an
analytical signaling model describing the residual CFO induced error after
data estimation. Once an error model is available, this information can then serve as a basis
for deriving reliability information in the form of e.g., bit log-likelihood
ratios (LLRs) and fed to a channel decoder
in the receiver. We will do so in Sec.\ref{sec:impact_cfo_ber}. For the calculation of the bit LLRs, we rely on the derivations
in \cite{Neshaastegaran19}, which assume as underlying model
\begin{align}
\est{r} = s\e^{j\theta} + w\label{equ:app_a000},
\end{align}
with $s\in\mathbb{A}$ denoting a transmitted symbol distorted by a residual phase
error given as $\theta\sim\mathcal{N}\left(0,\sigma_\theta^2\right)$ and
AWGN given as
$w\sim\mathcal{CN}\left(0,2\sigma_w^2\right)$. In
order to utilize the same LLR
calculation scheme, we have to bring the UW-OFDM signaling model into
the form of \eqref{equ:app_a000}. Based on this model, bit LLRs can
easily be derived by carrying out the equations in \cite{Neshaastegaran19},
which are left out here for reasons of compactness\footnote{We note that
  this work with $w\sim\mathcal{CN}\left(0,\sigma_w^2\right)$ and
  \cite{Neshaastegaran19} with $w\sim\mathcal{CN}\left(0,2\sigma_w^2\right)$) use slightly different definitions for a complex Gaussian distribution, which has to be accounted for in the LLR derivations.}.

Starting from \eqref{equ:cfo_rx031a} and compensating for the offset
in \eqref{equ:cfo_rx037}, an estimate of a single
data symbol follows as
\begin{align}
\kElemdataestl &=
\e^{j(\varphil-\varphilest)}\e^{-j\hat{\varphi}_\text{off}}\ve{e}_k^T\mf{\Lambda}_\text{stat}\mf{H}\ve{g}_k\kElemdatal\\
&\quad+\e^{-j\hat{\varphi}_\text{off}}\Delta_k+\e^{-j\hat{\varphi}_\text{off}}\e^{-j\varphilest}\ve{e}_k^T\noisef\label{equ:app001}\\
&=\alpha_k\kElemdatal\e^{j\theta_l}+\kElemnoisellr.\label{equ:app_a001}
\end{align}
We note that \eqref{equ:app_a001} coincides with \eqref{equ:app_a000},
whereas
$\alpha_k=\e^{-j\hat{\varphi}_\text{off}}\ve{e}_k^T\mf{\Lambda}_\text{stat}\mf{H}\ve{g}_k$
denotes a complex-valued scaling factor
such that $\alpha_k\kElemdataestl\equiv s$, and $\kElemnoisellr=
\e^{-j\hat{\varphi}_\text{off}}\Delta_k+\e^{-j\hat{\varphi}_\text{off}}\e^{-j\varphilest}\ve{e}_k^T\noisef$
is modelled as AWGN distributed according to $\kElemnoisellr\sim\mathcal{CN}\left(0,\sigma_{\kElemnoisellr}^2\right)$ with
\begin{align} \sigma_{\kElemnoisellr}^2=\E{\Delta_k\Delta_k^H}+\E{\ve{e}_k^T\noisef\noisef^H\ve{e}_k^*}.\label{equ:app_a002}
\end{align}
The assumption of a Gaussian distribution for $\kElemnoisellr$ seems justified, given that
$\noisef\sim\mathcal{CN}(\ve{0},\varnoisef\m{I})$,  see \eqref{equ:cfo_rx020}, and considering the law of large
numbers in case of $\Delta_k$, see \eqref{equ:cfo100}. Further,
$\theta_l=\varphil-\varphilest$ represents the residual phase error with an assumed distribution of
$\theta_l\sim\mathcal{N}\left(0,\sigma_{\theta}^2\right)$. In order to verify
this assumption, we have to investigate the estimate $\varphilest$ in detail
and derive $\sigma_{\theta}^2$ accordingly.

Throughout this work, we utilize a CPE estimation
approach \cite{Hofbauer20} given as
\begin{align}
\varphilest&= \text{arg}\left(\ve{p}^H\m{W}_p\pestl\right),\label{equ:app_a003}
\end{align}
where $\m{W}_p=\diag{\ve{w}_p}$, $\ve{w}_p\in\mathbb{{R}^+}^{N_p\times 1}$, forms a diagonal weighting
matrix and $\pestl\in\mathbb{C}^{N_p\times 1}$ incorporates the estimated pilot symbols from the $l$th OFDM symbol. According to \cite{Hofbauer20},
the $m$th estimated pilot symbol is given as
\begin{align}
\mElempestl&=\mel{\m{E}_p}_{m,*}\lambdafHl\left(\m{G}_p\ve{p} + \m{B}^T\vef{x}_u\right) + \mElemNNoisePil\nonumber\\
&=
\ve{e}^T_{m}\e^{\varphil}\mf{\Lambda}_{h,\text{stat}}\left(\ve{g}_{m}\mElemp+
\m{B}^T\vef{x}_u\right) + \mElemNNoisePil,\label{equ:app_a004}
\end{align}
with $\ve{e}^T_{m}=\mel{\m{E}_p}_{m,*}$, $\m{E}_p\in\mathbb{N}^{N_p\times
  (N-N_z)}$ denoting a pilot subcarrier selection matrix,
$\ve{g}_{m}=\mel{\m{G}_p}_{*,m}$, 
$\mf{\Lambda}_{h,\text{stat}}=\mf{H}^{-1}\mf{\Lambda}_\text{stat}\mf{H}$, and
$\mElemNNoisePil$ denoting an additive noise term detailed further in the subsequent paragraphs.
Inserting \eqref{equ:app_a004} into \eqref{equ:app_a003} together with
introducing the notation
\begin{align}
a_p\e^{j\varphi_p}=\sum_{m=0}^{N_p-1}\mElemwp\ve{e}^T_{m}\mf{\Lambda}_{h,\text{stat}}\left(\ve{g}_{m}|\mElemp|^2
+ \m{B}^T\vef{x}_u\mElemp^H\right)\label{equ:app_a005}
\end{align}
yields
\begin{align}
  \varphilest&= \text{arg}\left(\ve{p}^H\m{W}_p\pestl\right)\label{equ:app_a006}\\ &=\text{arg}\left(\sum_{m=0}^{N_p-1}\mElemp^H\mElemwp\mElempestl\right)\label{equ:app_a007}\\
&=\text{arg}\Biggl(\e^{j\varphil}\e^{j\varphi_p}+
  \frac{1}{a_p}\sum_{m=0}^{N_p-1}\mElemNNoisePil\mElemwp\mElemp^H\Biggr)\label{equ:app_a008}\\
&= \varphil + \varphi_p + \delta_l,\label{equ:app_a009}
\end{align}
whereas $\delta_l$ represents an additive deviation approximated by
\begin{equation}
\delta_l\approx f\left(\frac{1}{a_p}\sum_{k=0}^{N_p-1}\mElemNNoisePil\mElemwp\mElemp^H\right),\label{equ:app_a010}
\end{equation}
with some proper function $f(\cdot)$ modelling this approximation. Following
\cite{Norifumi13}, it holds that for any $\alpha\in[0,2\pi[$, $\beta\in\mathbb{R}$, and
$n\sim\mathcal{CN}\left(0,\sigma_{{n}}^2\right)$, an estimate $\est{\alpha}$ can be well approximated as
\begin{align}
\est{\alpha}=\text{arg}\left(\beta\e^{j\alpha}+n\right)\approx \alpha + n_{\alpha}\label{equ:app_a011}
\end{align}
with $n_{\alpha}\sim\mathcal{N}\left(0,\frac{1}{2}\sigma_n^2\right)$, given
that $|\beta|^2\gg\sigma_n^2$. Applying the approximation in \eqref{equ:app_a011} on
\eqref{equ:app_a009} and \eqref{equ:app_a010} requires further elaboration of
$\mElemNNoisePil$. As detailed in
\cite{Hofbauer20}, $\mElemNNoisePil$ consists of the three additive terms
\begin{align}
\mElemNNoisePil=\mElempicil + \mElemdataicil +\mElemNoisePil,\label{equ:app_a012}
\end{align}
which denote the $m$th element of the respective vectors
$\piloticil\in\mathbb{C}^{N_p\times 1}$, $\dataicil\in\mathbb{C}^{N_p\times 1}$
and $\noisefpil\in\mathbb{C}^{N_p\times 1}$ defined as
\begin{align}
\piloticil&=\m{E}_p\mf{H}^{-1}\lambdafl\mf{H}\m{G}_p\ve{p}
-\ve{p}\label{equ:app_a013},\\
\dataicil&=\m{E}_p\mf{H}^{-1}\lambdafl\mf{H}\m{G}_d\datal\label{equ:app_a014}, \\
\noisefpil&=\m{E}_p\mf{H}^{-1}\m{B}^T\noisef.\label{equ:app_a015}
\end{align}
In this case, $\piloticil$ denotes a constant offset due to ICI induced by the other pilot
symbols, $\dataicil$ a data induced ICI and $\noisefpil$ AWGN. With a mean
$\mu$ due to the pilot subcarrier induced ICI given as $\mu
=\frac{1}{a_p}\sum_{m=0}^{N_p-1}\mElemwp\mElemp^H\mElempicil$, under the
assumption of $\E{\mElemdataicil\mElemNoisePil^H}=0$ and with
$\sigma_{d_{\ici,m}}^2 = \E{\mElemdataicil\mElemdataicil^H}$ and $\sigma_{v_m''}^2 = \E{\mElemNoisePil\mElemNoisePil^H}$, we obtain
\begin{align}
\sigma_{{n_{\alpha}}}^2 &=
\frac{1}{2}\E{\left(n-\mu\right)\left(n-\mu\right)^H} \label{equ:app_a016}\\
&\equiv \sigma_\theta^2 \label{equ:app_a017}\\
&=\frac{1}{2|a_p|^2}\sum_{m=0}^{N_p-1}|\mElemwp|^2|\mElemp|^2\left(\sigma_{d_{\ici,m}}^2
+ \sigma_{v_m''}^2\right).\label{equ:app_a018}
\end{align}
The Gaussian distribution assumption of \eqref{equ:app_a018} is justified based on the same
arguments as for $\kElemnoisellr$, cf.~\eqref{equ:app_a002}. We note that \eqref{equ:app_a011} assumes $\mu=\E{n}=0$. Although this is not
entirely fulfilled for the UW-OFDM signaling model due to $\mElempicil$, we further note that pilot symbols
are usually approximately uniformly distributed over the entire subcarrier set to
optimize estimation performance of system parameters \cite{Cai04}. Consequently, the pilot symbols are several subcarriers apart
from each other, resulting in a rather minor ICI due to
$\piloticil$, which therefore justifies an assumption of $\mu\approx 0$.

As a last step, we still have to incorporate the constant phase offset
$\varphi_p$ within $\varphilest$ from \eqref{equ:app_a009} into our model by introducing
\begin{align}
\alpha_k'=\alpha_k\e^{j\varphi_p},
\end{align}
thus yielding the final signal model
\begin{align}
\kElemdataestl =\alpha'_k\kElemdatal\e^{-j\theta_l}+\kElemnoisellr,\label{equ:app_a019}
\end{align}
whereas $\theta\sim\mathcal{N}\left(0,\sigma_\theta^2\right)$ with $\sigma_\theta^2$
given in \eqref{equ:app_a018} as well as $\kElemnoisellr\sim\mathcal{CN}\left(0,\sigma_{\kElemnoisellr}^2\right)$
with $\sigma_{\kElemnoisellr}^2$ defined in \eqref{equ:app_a002}. We note that
for the advanced compensation methods tackling ICI impairments as well, see Sec.~\ref{sec:advanced_cfo_comp}, we
assume $\E{\Delta_k\Delta_k^H}\stackrel{!}{=}0$ within
$\sigma_{\kElemnoisellr}^2$. Based on this model, bit LLRs can
easily be derived following the equations in \cite{Neshaastegaran19}, which
concludes the derivations for the underlying signaling model.

\section{Bit Error Ratio Simulations}
\label{sec:impact_cfo_ber}

\begin{figure*}[thb]
\centering
\begin{tikzpicture}[x=1mm, y=1mm,font=\small] 
  \def\rectW{16}
  \def\rectH{14}
  \def\ArrowL{7}
  \def\ArrowLongL{10}
  \def\ArrowShortL{5}
  \def\ArrowSShortL{3}
    \def\OffsetOFDMSymb{-9}
  
 \node[align=center] (nBinaryInput) at (0,0) {binary \\ data input}; 
 \node[draw, rectangle, minimum width=\rectW, minimum height=\rectH,
   align=center,right=\ArrowShortL of nBinaryInput](nChannelCoding) {Channel \\ coding};
 \node[draw, rectangle, minimum width=\rectW, minimum height=\rectH,
   align=center, right=\ArrowShortL of nChannelCoding](nInterleaving) {Inter- \\ leaving};
 \node[draw, rectangle, minimum width=\rectW, minimum height=\rectH,
   align=center, right=\ArrowShortL of nInterleaving](nQAMMapping) {QAM \\ mapping};
 \node[draw, rectangle, minimum width=\rectW, minimum height=55,
   align=center, below right=\OffsetOFDMSymb and \ArrowLongL of nQAMMapping](nOFDMSymbGen) {OFDM \\symbol\\
   generation\\ $\m{B}$, $\m{G}_d$, $\m{G}_p$};

 \node[draw, rectangle, minimum width=\rectW, minimum height=\rectH,
   align=center, right=\ArrowL of nOFDMSymbGen](nIDFT) {IDFT};
 \node[draw, rectangle, minimum width=\rectW, minimum height=\rectH,
   align=center, below right=\ArrowSShortL of nIDFT](nChannel) {Channel
   \\[1mm] $\lambdaNfl$, $\mf{H}'$,$\noiseffull$};
 
 \draw[->] (nBinaryInput) -- (nChannelCoding);
 \draw[->] (nChannelCoding) -- (nInterleaving);
 \draw[->] (nInterleaving) -- (nQAMMapping);
 \draw[->] (nOFDMSymbGen) -- (nIDFT) node[midway, above] {$\Xfl$};
 \draw[->] (nIDFT) -| (nChannel);

 \draw[->] (nQAMMapping.east) -- (nOFDMSymbGen.west|- 0,0) node[midway, above]
      {$\datal$};

 \node (nPilot) at ([xshift=10] nQAMMapping.east|- 0,-7) {$\ve{p}$};
 \node (nXu) at ([xshift=-11, yshift=-11] nPilot) {$\vef{x}_u = \mathbf{0}$};

 \draw[->] (nPilot.east) -- (nOFDMSymbGen.west|- 0,-7);
  \draw[->] (nPilot.east|- 0,-11) -- (nOFDMSymbGen.west|- 0,-11);


 \node[draw, rectangle, minimum width=\rectW, minimum height=\rectH,
   align=center, below left=\ArrowSShortL of nChannel](nDFT) {DFT};
 \node[draw, rectangle, minimum width=\rectW, minimum height=\rectH,
   align=center, left=\ArrowL of nDFT](nPayloadExtraction) {$\m{B}^T$};
 \node[draw, rectangle, minimum width=\rectW, minimum height=\rectH,
   align=center, left=\ArrowL of nPayloadExtraction](nXoffCompensation)
      {$-\ve{x}_\text{off}$};
 \node[draw, rectangle, minimum width=\rectW, minimum height=\rectH,
   align=center, left=\ArrowL of nXoffCompensation](nCFOCompensation)
      {CFO comp.};
 \node[draw, rectangle, minimum width=\rectW, minimum height=\rectH,
   align=center, below left=\ArrowL of nPayloadExtraction](nCFOEstimation)
      {CFO estimation};
      
 \node[draw, rectangle, minimum width=\rectW, minimum height=\rectH,
   align=center, left=\ArrowL of nCFOCompensation](nDataEstimation) {Data est.
   \\ $\m{E}_\lmmse$}; 
 \node[draw, rectangle, minimum width=\rectW, minimum height=\rectH,
   align=center, left=\ArrowL of nDataEstimation](nQAMDemapping) {QAM \\ demap.};
 \node[draw, rectangle, minimum width=\rectW, minimum height=\rectH,
   align=center, left=\ArrowShortL of nQAMDemapping](nDeinterleaving) {Deinter- \\ leaving}; 
  \node[draw, rectangle, minimum width=\rectW, minimum height=\rectH,
   align=center, left=\ArrowShortL of nDeinterleaving](nViterbi){Viterbi \\ decoding};
 \node[align=center, left=\ArrowShortL of nViterbi] (nBinaryOutput) {binary \\ data \\output}; 

  \draw[<-] (nBinaryOutput) -- (nViterbi);  
 \draw[<-] (nViterbi) -- (nDeinterleaving);
 \draw[<-] (nDeinterleaving) -- (nQAMDemapping);
 \draw[<-] (nQAMDemapping) -- (nDataEstimation) node[midway, above] {$\dataestl$};
 \draw[<-] (nDataEstimation) -- (nCFOCompensation);
  \draw[<-] (nCFOCompensation) -- (nXoffCompensation); 

 \draw[<-] (nXoffCompensation) -- (nPayloadExtraction) node[midway, above]
      {$\Yfdownl$};
 \draw[<-] (nPayloadExtraction) -- (nDFT) node[midway, above]  {$\Yfrxl$};
 \draw[<-] (nDFT) -| (nChannel);

 \node (nBranchArrow) at ($(nDFT.west) + (-\ArrowL/3, +1.75)$) {};

 
 \draw[<-{Circle[]}] (nCFOEstimation) -| (nBranchArrow);
 \draw[<-] (nCFOCompensation.south) |- (nCFOEstimation.west) node[pos=0.35,
   left]{$\varphilest$, $\est{\epsilon}$};
\end{tikzpicture}

\caption{Block diagram of the considered UW-OFDM transceiver system.}
\label{fig:block_diagram}
\end{figure*}

In the following, we will present BER simulations for coded as well as uncoded
transmission in a frequency selective environment. So far, CFO effects
have been considered in this work in isolation to separate its influence on UW-OFDM
from other degrading effects. As such, we intentionally relied on the
simplifying assumption of $\varnoisef=0$. In this section, we investigate the CFO impairments w.r.t.\ a full UW-OFDM communication system and thus consider $\varnoisef\neq 0$. We note that
this work substantially extends the BER analysis conducted in previous works, as
up till now BER assessments have either been restricted to considerations without CFO
impairments \cite{Huemer12_1}, or the receiver did not incorporate accurate information on the residual CFO
impairments in the decoding process \cite{Hofbauer16}.


\subsection{Simulation setup}\label{sec:setup}

We generate UW-OFDM packets with $L = 200$ UW-OFDM symbols and process them as
depicted in Fig.~\ref{fig:block_diagram}. We apply a rate $r=1/2$ convolutional code with constraint length
$7$ and generator polynomial $(133;171)_8$ as channel encoder, as well as a
code of rate $r=3/4$ derived from the $r=1/2$ code according to the puncturing pattern $\left(\begin{smallmatrix}1 & 1 & 0\\1 & 0 &
  1 \end{smallmatrix}\right)$. The encoded bits are
interleaved within one UW-OFDM packet. We note that the interleaver length is
different compared to an interleaver restricted
to one OFDM symbol as utilized in some of our previous works \cite{Huemer10_1,Huemer12_1,Huemer11_1,Hofbauer16,
  Hofbauer16_1}, which therefore may  impede a direct comparison of the BER
results in some cases. However, a change of the interleaver length is necessary to prevent statistical dependencies among the bits
within an OFDM symbol $l$, which are otherwise present due to a common error
introduced by the estimate $\varphilest$, resulting then in a degradation of the decoding performance at the receiver. QPSK and QAM16 serve as modulation
alphabet and $\m{G}'_d$ as well as $\m{G}''_d$ are applied as generator matrices. Soft decision Viterbi decoding is applied at the receiver, with the
reliability information provided in form of LLRs
derived from the signaling models for $\dataestl$ given in \eqref{equ:cfo_rx040}
and \eqref{equ:cfo_comp005} for simple CPE and advanced CFO compensation,
respectively, and with the detailed model derivations presented in
Sec.~\ref{sec:error_model}.

We scale the utilized UW-OFDM generator matrices $\m{G }'_d$ and $\m{G }''_d$ such that
$\m{G}_d^{'H}\m{G}'_d=\alpha\m{I}$ and $\m{G}_d^{''H}\m{G}''_d=\alpha\m{I}$ with $\alpha=N'_{d}/N_d$, whereas
$N'_{d}$ corresponds to the number of CP-OFDM data subcarriers. This scaling will provide a fair comparison with CP-OFDM, as it ensures
for both the same data induced mean power per non-pilot subcarrier, which
directly effects the level of data induced ICI disturbances, see
\eqref{equ:cfo100}. In fact, it is even slightly in favor of CP-OFDM, as the
spreading by $\m{G}_p\ve{p}$ adds up to the
total mean power per non-pilot subcarrier. We use $\m{G}_{d,\cpofdm}=\m{B}_p\m{I}$ and $\m{G}_{p,\cp}=\m{P}_{p}\begin{bmatrix}\m{I} &
  \m{0}^T \end{bmatrix}^T$ to model CP-OFDM. Furthermore, Tab.~\ref{tab:setups} presents the relevant setup parameters of the
investigated UW-OFDM and CP-OFDM systems.
\begin{table}[t!]
\caption{Main parameters of the utilized UW-OFDM and
  CP-OFDM setup.}
\begin{center}
\begin{scriptsize}
\begin{tabular}{lc|rr} \hline
&& UW-OFDM & CP-OFDM \\ \hline\hline
DFT size & $N$ & 64 & 64 \\ \hline
data subcarriers & $N_d$, $N'_d$ & 32 & 48 \\ \hline
zero subcarriers & $N_z$ & 12 & 12 \\ \hline
pilot subcarriers & $N_p$ & 4 & 4 \\ \hline
red. subcarriers & $N_r$ & 16 & -\\ \hline
guard interval samples & $N_g,N_u$ & 16 & 16\\ \hline
 & & & \\
zero subcarrier indices & $\mathcal{I}_z$ & \{0,27,28,\dots,37\} &\{0,27,28,\dots,37\}\\
 & & & \\
pilot subcarrier indices & $\mathcal{I}_p$ & \{7,21,43,57\} & \{7,21,43,57\}\\
 & & & \\ \hline
DFT length & $T_\text{DFT}$ &  3.2\,$\mu$s & 3.2\,$\mu$s\\ \hline
guard interval length & $T_\text{GI}$ & 0.8\,$\mu$s & 0.8\,$\mu$s\\ \hline
OFDM symbol lengh & $T_\text{OFDM}$ & 3.2\,$\mu$s & 4\,$\mu$s \\ \hline
subcarrier spacing & $\Delta_f$ & 312.5\,kHz& 312.5\,kHz\\ \hline
\end{tabular}
\end{scriptsize}
\label{tab:setups}
\end{center}
\end{table}

BER results are obtained by averaging over $10^4$ independent channel
realizations, with the channel impulse responses normalized to unit energy and
following an exponentially decaying power delay profile \cite{Fak97} with a
given channel delay spread of $\delayspread=100$\,ns. We have carried out
simulations for several CFO values in the range $0 \leq \epsilon \leq
0.1$, but we only show results for 0 and 0.1 to enhance
clarity in the figures. All other results fall within the corridor spanned by those
boarder values. Additionally, BER results for the case without CFO serve as principle
performance bounds in the following figures. Since the pilot subcarriers
are utilized for estimating $\varphil$, we choose $\ve{x}_u=\ve{0}$. Except for the bounds, all results
presented in Figs.~\ref{fig:BER_100ns_cfo_unc}--\ref{fig:BER_100ns_cfo_r34_qam16_g1g2} always incorporate a CFO compensation by multiplying the
$l$th OFDM symbol with $\e^{-j(\varphilest + \est{\varphi}_\text{off})}$, which
is thus omitted in the legend. 

The estimate $\varphilest$ in 
\eqref{equ:cfo_varphilest} follows from $\varphilest = \text{arg}\left(\ve{p}^H\m{W}_p\pestl\right)$ with
$\m{W}_p=|\mf{H}_p|^2$ and $\mf{H}_p\in\mathbb{C}^{N_p\times N_p}$ denoting a
diagonal matrix with the channel frequency response coefficients corresponding to the pilot
subcarriers on its main diagonal. For all systems we apply a simple pilot subcarrier
extraction matrix $\m{E}_p=\begin{bmatrix}\m{0} &\m{I} \end{bmatrix}\m{P}_p^T$ to estimate $\pestl$. The phase offset $\est{\varphi}_\text{off}$ is
estimated based on the model stated in \eqref{equ:est_varphi_off}, with the
details given in \cite{Hofbauer20}. We assume perfect knowledge of
the channel and evaluate performance differences among systems at a BER of $10^{-6}$.

\subsection{Results}
The MSE analyses conducted in Sec.~\ref{sec:cpe_correction} showed
a very similar robustness of $\m{G}'_d$ and $\m{G}''_d$ against CFO and thus an
almost identical performance in terms of this error metric. However, the same
findings cannot not be expected per se for BER considerations of a complete communication transceiver
chain. As laid out in e.g., \cite{Huemer12_1}, UW-OFDM systems with different generator matrices show a quite different
BER behavior depending on the transceiver chain, e.g., the properties of the
wireless channel or the coding rate of the applied channel code. As such,
individual investigations of $\m{G}'_d$ and $\m{G}''_d$ are necessary in the following.
In Fig.~\ref{fig:BER_100ns_cfo_unc}, we compare UW-OFDM
($\m{G}'_d$ and $\m{G}''_d$) against
CP-OFDM ($\m{G}_{d,\cpofdm}$) for $\epsilon=0$ and $\epsilon=0.1$ in case of uncoded
transmission. Both UW-OFDM systems significantly outperform
CP-OFDM in the high $E_b/N_0$ regime, with superior performance of
$\m{G}''_d$. While all three systems experience a saturating BER behavior
for $\epsilon=0.1$ eventually, this happens for $\m{G}''_d$ not before the very low BER
regime. For all systems, considerations without CFO slightly deviate from $\epsilon=0$, but this is only noticeable in lower $E_b/N_0$ regimes (see zoom in
Fig.~\ref{fig:BER_100ns_cfo_unc_zoom}). The reason for this difference is due to the
CFO estimation algorithm incorrectly detecting $\hat{\epsilon}\neq 0$ (due to $\varphilest\neq 0$) instead of $\epsilon= 0$, leading to a
performance gap to curves without CFO, with the latter representing the performance
bound of $\epsilon=0$ and error free detection.
\begin{figure}[!tbh]
\centering
\scalebox{0.66}{
%
\definecolor{mycolor1}{rgb}{1.00000,0.50196,0.00000}%
\definecolor{mycolor2}{rgb}{0.12941,0.44314,0.75686}%
\begin{tikzpicture}

\begin{axis}[%
width=4.521in,
height=3.566in,
at={(0.758in,0.481in)},
scale only axis,
xmin=0,
xmax=35,
xlabel style={font=\color{white!15!black}},
xlabel={$E_b/N_0$ (dB)},
ymin=-6,
ymax=0,
ylabel style={font=\color{white!15!black}},
ylabel={$\textrm{log}_{10}(\text{BER})$},
axis background/.style={fill=white},
xmajorgrids,
ymajorgrids,
legend style={at={(0.03,0.03)}, anchor=south west, legend cell align=left, align=left, draw=white!15!black}
]
\addplot [color=black, dotted, line width=1.0pt, mark=square*, mark options={solid, fill=black, black}]
  table[row sep=crcr]{%
0	-0.6378\\
1	-0.6837\\
2	-0.7352\\
3	-0.7923\\
4	-0.8551\\
5	-0.924\\
6	-0.9981\\
7	-1.077\\
8	-1.1598\\
9	-1.2463\\
10	-1.3356\\
11	-1.4273\\
12	-1.5206\\
13	-1.6155\\
14	-1.7107\\
15	-1.8073\\
16	-1.9041\\
17	-2.0026\\
18	-2.0994\\
19	-2.1978\\
20	-2.2963\\
21	-2.3954\\
22	-2.4915\\
23	-2.5919\\
24	-2.6906\\
25	-2.7881\\
26	-2.8857\\
27	-2.9837\\
28	-3.0803\\
29	-3.1758\\
30	-3.2717\\
31	-3.37\\
32	-3.4688\\
33	-3.5675\\
34	-3.6615\\
35	-3.7623\\
};
\addlegendentry{$\m{G}_{d,\cpofdm}$, w/o CFO}

\addplot [color=black, line width=1.0pt, mark=square*, mark options={solid, fill=black, black}]
  table[row sep=crcr]{%
0	-0.574776118895735\\
1	-0.621892568138268\\
2	-0.674740110683919\\
3	-0.733516008616905\\
4	-0.798376568897065\\
5	-0.869852857375277\\
6	-0.947814643459898\\
7	-1.03145539943164\\
8	-1.11968229596372\\
9	-1.21210746071679\\
10	-1.30690282420876\\
11	-1.40384687436234\\
12	-1.50152790836889\\
13	-1.5998597203253\\
14	-1.69874420729395\\
15	-1.79747152595268\\
16	-1.89654105161379\\
17	-1.99523452567103\\
18	-2.09480528707427\\
19	-2.19365995184153\\
20	-2.29282473592349\\
21	-2.39278575972671\\
22	-2.49159163671852\\
23	-2.59044461939385\\
24	-2.68844369075951\\
25	-2.78626509472089\\
26	-2.88573661265685\\
27	-2.98236461639225\\
28	-3.07953153720376\\
29	-3.17658011538109\\
30	-3.27067909626908\\
31	-3.36612143083422\\
32	-3.46584824820831\\
33	-3.56688883873123\\
34	-3.65873038694156\\
35	-3.76528767365747\\
};
\addlegendentry{$\m{G}_{d,\cpofdm}$, $\epsilon=0$}

\addplot [color=black, line width=1.0pt, mark=triangle*, mark options={solid, rotate=270, fill=black, black}]
  table[row sep=crcr]{%
0	-0.563684321278648\\
1	-0.608191867341884\\
2	-0.657938528043467\\
3	-0.712736819794458\\
4	-0.772910098596732\\
5	-0.838750977054096\\
6	-0.91014813041334\\
7	-0.986332555113117\\
8	-1.06690971030622\\
9	-1.15118030497474\\
10	-1.23811881192067\\
11	-1.32682793828804\\
12	-1.41605878371775\\
13	-1.50513485065893\\
14	-1.59347410113493\\
15	-1.68025888936554\\
16	-1.76426789380506\\
17	-1.84625339636587\\
18	-1.9248506775941\\
19	-1.99948074053423\\
20	-2.0705007159525\\
21	-2.13639990706503\\
22	-2.19495991788227\\
23	-2.24986339908644\\
24	-2.29847248910405\\
25	-2.34196225614812\\
26	-2.38059467580708\\
27	-2.41286162143733\\
28	-2.44093915799948\\
29	-2.46419443966618\\
30	-2.48425651384753\\
31	-2.50010466745364\\
32	-2.51264211608836\\
33	-2.52466028740281\\
34	-2.53322348481557\\
35	-2.5404598822383\\
};
\addlegendentry{$\m{G}_{d,\cpofdm}$, $\epsilon=0.1$}

\addplot [color=mycolor1, dotted, line width=1.0pt, mark=square*, mark options={solid, fill=mycolor1, mycolor1}]
  table[row sep=crcr]{%
0	-0.6812\\
1	-0.7332\\
2	-0.7918\\
3	-0.857\\
4	-0.9289\\
5	-1.0075\\
6	-1.0931\\
7	-1.184\\
8	-1.2809\\
9	-1.3826\\
10	-1.4895\\
11	-1.6003\\
12	-1.7152\\
13	-1.8351\\
14	-1.9593\\
15	-2.0886\\
16	-2.2224\\
17	-2.3633\\
18	-2.5102\\
19	-2.6644\\
20	-2.8269\\
21	-2.9982\\
22	-3.1741\\
23	-3.3573\\
24	-3.5495\\
25	-3.7518\\
26	-3.9621\\
27	-4.1723\\
28	-4.389\\
29	-4.616\\
30	-4.8517\\
31	-5.124\\
32	-5.3438\\
33	-5.6144\\
34	-5.8592\\
35	-6.153\\
};
\addlegendentry{$\m{G}_d'$, w/o CFO}

\addplot [color=mycolor1, line width=1.0pt, mark=square*, mark options={solid, fill=mycolor1, mycolor1}]
  table[row sep=crcr]{%
0	-0.589255326753049\\
1	-0.641527424577637\\
2	-0.700259225915411\\
3	-0.766603530401824\\
4	-0.840293540571901\\
5	-0.922268678502914\\
6	-1.01207749120848\\
7	-1.10955819620606\\
8	-1.21446930574938\\
9	-1.32474702046642\\
10	-1.44012309913129\\
11	-1.55886895931195\\
12	-1.68122929444155\\
13	-1.80620451231533\\
14	-1.93472584356709\\
15	-2.06843606367549\\
16	-2.20641694963869\\
17	-2.34945116800221\\
18	-2.49808425372545\\
19	-2.65427869113068\\
20	-2.8162431830696\\
21	-2.98760111206965\\
22	-3.16354236401514\\
23	-3.35407138997977\\
24	-3.54781810307498\\
25	-3.74502919305534\\
26	-3.95486046133514\\
27	-4.17029428870261\\
28	-4.39003894281556\\
29	-4.62763465947288\\
30	-4.85557174919966\\
31	-5.11421487121653\\
32	-5.32689265750772\\
33	-5.59532660866899\\
34	-5.9253663817031\\
35	-6.18813187727179\\
};
\addlegendentry{$\m{G}_d'$, $\epsilon=0$}

\addplot [color=mycolor1, line width=1.0pt, mark=x, mark options={solid, fill=mycolor1, mycolor1}]
  table[row sep=crcr]{%
0	-0.580690962558614\\
1	-0.630967370221224\\
2	-0.687644039118388\\
3	-0.751496189568902\\
4	-0.821812839078318\\
5	-0.899949019066852\\
6	-0.985131743567159\\
7	-1.07802991167134\\
8	-1.17722555209174\\
9	-1.28276819889033\\
10	-1.39260881874513\\
11	-1.50680049826361\\
12	-1.62476675074506\\
13	-1.74455298278882\\
14	-1.86711597071745\\
15	-1.99314703578687\\
16	-2.12222340215061\\
17	-2.25423679513748\\
18	-2.39021306200119\\
19	-2.5288928758073\\
20	-2.6712828281467\\
21	-2.81928737044597\\
22	-2.97034043650497\\
23	-3.12357466300159\\
24	-3.28136209647658\\
25	-3.43392174480632\\
26	-3.59519300015174\\
27	-3.75271736905843\\
28	-3.90987426651781\\
29	-4.06110518240183\\
30	-4.19344908323554\\
31	-4.34498168536139\\
32	-4.46139485135123\\
33	-4.56215038495387\\
34	-4.67988618329062\\
35	-4.74813374358861\\
};
\addlegendentry{$\m{G}_d'$, $\epsilon=0.1$}

\addplot [color=mycolor2, dotted, line width=1.0pt, mark=o, mark options={solid, mycolor2}]
  table[row sep=crcr]{%
0	-0.6879933650809\\
1	-0.740257234901373\\
2	-0.799690377781048\\
3	-0.867540834243622\\
4	-0.945665859272405\\
5	-1.03481300312611\\
6	-1.13743484118894\\
7	-1.25448868023778\\
8	-1.38859436607943\\
9	-1.54056993250286\\
10	-1.71224549376719\\
11	-1.90276520190745\\
12	-2.11180470490431\\
13	-2.33945318866666\\
14	-2.58025845857014\\
15	-2.83485540904595\\
16	-3.10044024140239\\
17	-3.37678302438182\\
18	-3.65636065624018\\
19	-3.94175789681594\\
20	-4.24216524795477\\
21	-4.52118558726089\\
22	-4.82960075534378\\
23	-5.14626377391404\\
24	-5.4608062434248\\
25	-5.72872\\
26	-6.00964\\
};
\addlegendentry{$\m{G}_d''$, w/o CFO}

\addplot [color=mycolor2, line width=1.0pt, mark=o, mark options={solid, mycolor2}]
  table[row sep=crcr]{%
0	-0.596849238583285\\
1	-0.649317011407837\\
2	-0.708657759504374\\
3	-0.774981577260283\\
4	-0.851060063237548\\
5	-0.937099051041248\\
6	-1.03557686967257\\
7	-1.14858183088494\\
8	-1.2778248787674\\
9	-1.42663017763327\\
10	-1.59562457005856\\
11	-1.78540096673586\\
12	-1.99831765723072\\
13	-2.22992963932003\\
14	-2.48078644746107\\
15	-2.74650865847794\\
16	-3.02045873840581\\
17	-3.30935407105953\\
18	-3.6026439669846\\
19	-3.89750733752752\\
20	-4.2048262852154\\
21	-4.50222834004044\\
22	-4.7952443092795\\
23	-5.08726328796903\\
24	-5.39456026802066\\
25	-5.69534\\
26	-5.97702\\
};
\addlegendentry{$\m{G}_d''$, $\epsilon=0$}

\addplot [color=mycolor2, line width=1.0pt, mark=triangle, mark options={solid, rotate=90, mycolor2}]
  table[row sep=crcr]{%
0	-0.589036017988203\\
1	-0.639891865381391\\
2	-0.6967291189789\\
3	-0.760471231371647\\
4	-0.832770266962531\\
5	-0.913871033052149\\
6	-1.00572657598988\\
7	-1.1097165103451\\
8	-1.22734188206219\\
9	-1.36102726686086\\
10	-1.51030644720956\\
11	-1.67596348083188\\
12	-1.85947108925193\\
13	-2.05915368610843\\
14	-2.27378999481041\\
15	-2.50090170976107\\
16	-2.74188113042536\\
17	-2.98957415225837\\
18	-3.24555609943396\\
19	-3.50600855804851\\
20	-3.77165178158201\\
21	-4.0411077728811\\
22	-4.29920166973747\\
23	-4.56675535597646\\
24	-4.79971393173466\\
25	-5.12133461233947\\
26	-5.34082\\
27	-5.56922\\
28	-5.79552\\
29	-6.0505\\
30	-6.24214\\
31	-6.50514997831991\\
32	-6.64481197174891\\
33	-6.85193746454456\\
34	-7.02802872360024\\
35	-6.99326661734103\\
};
\addlegendentry{$\m{G}_d''$, $\epsilon=0.1$}

\end{axis}
\end{tikzpicture}%
}
\caption{BER results for uncoded transmission and QPSK.}
\label{fig:BER_100ns_cfo_unc}
\end{figure}
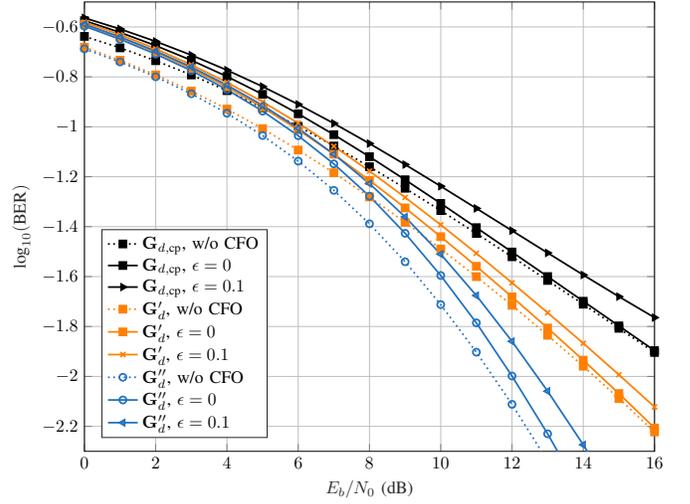
\begin{figure}[!tbh]
\centering
\scalebox{0.66}{
%
\definecolor{mycolor1}{rgb}{1.00000,0.50196,0.00000}%
\definecolor{mycolor2}{rgb}{0.12941,0.44314,0.75686}%
\begin{tikzpicture}

\begin{axis}[%
width=4.521in,
height=3.566in,
at={(0.758in,0.481in)},
scale only axis,
xmin=0,
xmax=16,
xlabel style={font=\color{white!15!black}},
xlabel={$E_b/N_0$ (dB)},
ymin=-2.3,
ymax=-0.5,
ylabel style={font=\color{white!15!black}},
ylabel={$\textrm{log}_{10}(\text{BER})$},
axis background/.style={fill=white},
xmajorgrids,
ymajorgrids,
legend style={at={(0.03,0.03)}, anchor=south west, legend cell align=left, align=left, draw=white!15!black}
]
\addplot [color=black, dotted, line width=1.0pt, mark=square*, mark options={solid, fill=black, black}]
  table[row sep=crcr]{%
0	-0.6378\\
1	-0.6837\\
2	-0.7352\\
3	-0.7923\\
4	-0.8551\\
5	-0.924\\
6	-0.9981\\
7	-1.077\\
8	-1.1598\\
9	-1.2463\\
10	-1.3356\\
11	-1.4273\\
12	-1.5206\\
13	-1.6155\\
14	-1.7107\\
15	-1.8073\\
16	-1.9041\\
17	-2.0026\\
18	-2.0994\\
19	-2.1978\\
20	-2.2963\\
21	-2.3954\\
22	-2.4915\\
23	-2.5919\\
24	-2.6906\\
25	-2.7881\\
26	-2.8857\\
27	-2.9837\\
28	-3.0803\\
29	-3.1758\\
30	-3.2717\\
31	-3.37\\
32	-3.4688\\
33	-3.5675\\
34	-3.6615\\
35	-3.7623\\
};
\addlegendentry{$\m{G}_{d,\cpofdm}$, w/o CFO}

\addplot [color=black, line width=1.0pt, mark=square*, mark options={solid, fill=black, black}]
  table[row sep=crcr]{%
0	-0.574776118895735\\
1	-0.621892568138268\\
2	-0.674740110683919\\
3	-0.733516008616905\\
4	-0.798376568897065\\
5	-0.869852857375277\\
6	-0.947814643459898\\
7	-1.03145539943164\\
8	-1.11968229596372\\
9	-1.21210746071679\\
10	-1.30690282420876\\
11	-1.40384687436234\\
12	-1.50152790836889\\
13	-1.5998597203253\\
14	-1.69874420729395\\
15	-1.79747152595268\\
16	-1.89654105161379\\
17	-1.99523452567103\\
18	-2.09480528707427\\
19	-2.19365995184153\\
20	-2.29282473592349\\
21	-2.39278575972671\\
22	-2.49159163671852\\
23	-2.59044461939385\\
24	-2.68844369075951\\
25	-2.78626509472089\\
26	-2.88573661265685\\
27	-2.98236461639225\\
28	-3.07953153720376\\
29	-3.17658011538109\\
30	-3.27067909626908\\
31	-3.36612143083422\\
32	-3.46584824820831\\
33	-3.56688883873123\\
34	-3.65873038694156\\
35	-3.76528767365747\\
};
\addlegendentry{$\m{G}_{d,\cpofdm}$, $\epsilon=0$}

\addplot [color=black, line width=1.0pt, mark=triangle*, mark options={solid, rotate=270, fill=black, black}]
  table[row sep=crcr]{%
0	-0.563684321278648\\
1	-0.608191867341884\\
2	-0.657938528043467\\
3	-0.712736819794458\\
4	-0.772910098596732\\
5	-0.838750977054096\\
6	-0.91014813041334\\
7	-0.986332555113117\\
8	-1.06690971030622\\
9	-1.15118030497474\\
10	-1.23811881192067\\
11	-1.32682793828804\\
12	-1.41605878371775\\
13	-1.50513485065893\\
14	-1.59347410113493\\
15	-1.68025888936554\\
16	-1.76426789380506\\
17	-1.84625339636587\\
18	-1.9248506775941\\
19	-1.99948074053423\\
20	-2.0705007159525\\
21	-2.13639990706503\\
22	-2.19495991788227\\
23	-2.24986339908644\\
24	-2.29847248910405\\
25	-2.34196225614812\\
26	-2.38059467580708\\
27	-2.41286162143733\\
28	-2.44093915799948\\
29	-2.46419443966618\\
30	-2.48425651384753\\
31	-2.50010466745364\\
32	-2.51264211608836\\
33	-2.52466028740281\\
34	-2.53322348481557\\
35	-2.5404598822383\\
};
\addlegendentry{$\m{G}_{d,\cpofdm}$, $\epsilon=0.1$}

\addplot [color=mycolor1, dotted, line width=1.0pt, mark=square*, mark options={solid, fill=mycolor1, mycolor1}]
  table[row sep=crcr]{%
0	-0.6812\\
1	-0.7332\\
2	-0.7918\\
3	-0.857\\
4	-0.9289\\
5	-1.0075\\
6	-1.0931\\
7	-1.184\\
8	-1.2809\\
9	-1.3826\\
10	-1.4895\\
11	-1.6003\\
12	-1.7152\\
13	-1.8351\\
14	-1.9593\\
15	-2.0886\\
16	-2.2224\\
17	-2.3633\\
18	-2.5102\\
19	-2.6644\\
20	-2.8269\\
21	-2.9982\\
22	-3.1741\\
23	-3.3573\\
24	-3.5495\\
25	-3.7518\\
26	-3.9621\\
27	-4.1723\\
28	-4.389\\
29	-4.616\\
30	-4.8517\\
31	-5.124\\
32	-5.3438\\
33	-5.6144\\
34	-5.8592\\
35	-6.153\\
};
\addlegendentry{$\m{G}_d'$, w/o CFO}

\addplot [color=mycolor1, line width=1.0pt, mark=square*, mark options={solid, fill=mycolor1, mycolor1}]
  table[row sep=crcr]{%
0	-0.589255326753049\\
1	-0.641527424577637\\
2	-0.700259225915411\\
3	-0.766603530401824\\
4	-0.840293540571901\\
5	-0.922268678502914\\
6	-1.01207749120848\\
7	-1.10955819620606\\
8	-1.21446930574938\\
9	-1.32474702046642\\
10	-1.44012309913129\\
11	-1.55886895931195\\
12	-1.68122929444155\\
13	-1.80620451231533\\
14	-1.93472584356709\\
15	-2.06843606367549\\
16	-2.20641694963869\\
17	-2.34945116800221\\
18	-2.49808425372545\\
19	-2.65427869113068\\
20	-2.8162431830696\\
21	-2.98760111206965\\
22	-3.16354236401514\\
23	-3.35407138997977\\
24	-3.54781810307498\\
25	-3.74502919305534\\
26	-3.95486046133514\\
27	-4.17029428870261\\
28	-4.39003894281556\\
29	-4.62763465947288\\
30	-4.85557174919966\\
31	-5.11421487121653\\
32	-5.32689265750772\\
33	-5.59532660866899\\
34	-5.9253663817031\\
35	-6.18813187727179\\
};
\addlegendentry{$\m{G}_d'$, $\epsilon=0$}

\addplot [color=mycolor1, line width=1.0pt, mark=x, mark options={solid, fill=mycolor1, mycolor1}]
  table[row sep=crcr]{%
0	-0.580690962558614\\
1	-0.630967370221224\\
2	-0.687644039118388\\
3	-0.751496189568902\\
4	-0.821812839078318\\
5	-0.899949019066852\\
6	-0.985131743567159\\
7	-1.07802991167134\\
8	-1.17722555209174\\
9	-1.28276819889033\\
10	-1.39260881874513\\
11	-1.50680049826361\\
12	-1.62476675074506\\
13	-1.74455298278882\\
14	-1.86711597071745\\
15	-1.99314703578687\\
16	-2.12222340215061\\
17	-2.25423679513748\\
18	-2.39021306200119\\
19	-2.5288928758073\\
20	-2.6712828281467\\
21	-2.81928737044597\\
22	-2.97034043650497\\
23	-3.12357466300159\\
24	-3.28136209647658\\
25	-3.43392174480632\\
26	-3.59519300015174\\
27	-3.75271736905843\\
28	-3.90987426651781\\
29	-4.06110518240183\\
30	-4.19344908323554\\
31	-4.34498168536139\\
32	-4.46139485135123\\
33	-4.56215038495387\\
34	-4.67988618329062\\
35	-4.74813374358861\\
};
\addlegendentry{$\m{G}_d'$, $\epsilon=0.1$}

\addplot [color=mycolor2, dotted, line width=1.0pt, mark=o, mark options={solid, mycolor2}]
  table[row sep=crcr]{%
0	-0.6879933650809\\
1	-0.740257234901373\\
2	-0.799690377781048\\
3	-0.867540834243622\\
4	-0.945665859272405\\
5	-1.03481300312611\\
6	-1.13743484118894\\
7	-1.25448868023778\\
8	-1.38859436607943\\
9	-1.54056993250286\\
10	-1.71224549376719\\
11	-1.90276520190745\\
12	-2.11180470490431\\
13	-2.33945318866666\\
14	-2.58025845857014\\
15	-2.83485540904595\\
16	-3.10044024140239\\
17	-3.37678302438182\\
18	-3.65636065624018\\
19	-3.94175789681594\\
20	-4.24216524795477\\
21	-4.52118558726089\\
22	-4.82960075534378\\
23	-5.14626377391404\\
24	-5.4608062434248\\
25	-5.72339460366744\\
};
\addlegendentry{$\m{G}_d''$, w/o CFO}

\addplot [color=mycolor2, line width=1.0pt, mark=o, mark options={solid, mycolor2}]
  table[row sep=crcr]{%
0	-0.596849238583285\\
1	-0.649317011407837\\
2	-0.708657759504374\\
3	-0.774981577260283\\
4	-0.851060063237548\\
5	-0.937099051041248\\
6	-1.03557686967257\\
7	-1.14858183088494\\
8	-1.2778248787674\\
9	-1.42663017763327\\
10	-1.59562457005856\\
11	-1.78540096673586\\
12	-1.99831765723072\\
13	-2.22992963932003\\
14	-2.48078644746107\\
15	-2.74650865847794\\
16	-3.02045873840581\\
17	-3.30935407105953\\
18	-3.6026439669846\\
19	-3.89750733752752\\
20	-4.2048262852154\\
21	-4.50222834004044\\
22	-4.7952443092795\\
23	-5.08726328796903\\
24	-5.39456026802066\\
25	-5.74172198475697\\
};
\addlegendentry{$\m{G}_d''$, $\epsilon=0$}

\addplot [color=mycolor2, line width=1.0pt, mark=triangle, mark options={solid, rotate=90, mycolor2}]
  table[row sep=crcr]{%
0	-0.589036017988203\\
1	-0.639891865381391\\
2	-0.6967291189789\\
3	-0.760471231371647\\
4	-0.832770266962531\\
5	-0.913871033052149\\
6	-1.00572657598988\\
7	-1.1097165103451\\
8	-1.22734188206219\\
9	-1.36102726686086\\
10	-1.51030644720956\\
11	-1.67596348083188\\
12	-1.85947108925193\\
13	-2.05915368610843\\
14	-2.27378999481041\\
15	-2.50090170976107\\
16	-2.74188113042536\\
17	-2.98957415225837\\
18	-3.24555609943396\\
19	-3.50600855804851\\
20	-3.77165178158201\\
21	-4.0411077728811\\
22	-4.29920166973747\\
23	-4.56675535597646\\
24	-4.79971393173466\\
25	-5.12133461233947\\
26	-5.32330639037513\\
27	-5.55332694300399\\
28	-5.83074816547462\\
29	-6.02085013897312\\
30	-6.28113516694704\\
31	-6.50514997831991\\
32	-6.64481197174891\\
33	-6.85193746454456\\
34	-7.02802872360024\\
35	-6.99326661734103\\
};
\addlegendentry{$\m{G}_d''$, $\epsilon=0.1$}

\end{axis}
\end{tikzpicture}%
}
\caption{BER results for uncoded transmission and QPSK (zoom of Fig.~\ref{fig:BER_100ns_cfo_unc}).}
\label{fig:BER_100ns_cfo_unc_zoom}
\end{figure}
Fig.~\ref{fig:BER_100ns_cfo_r12_qpsk_g1g2} shows BER results for coded transmission for $r=1/2$. UW-OFDM with $\m{G}'_d$ outperforms CP-OFDM in all three considered scenarios, namely by
$1.6$\,dB in case without CFO, $0.3$\,dB for $\epsilon=0$ and $1.0$\,dB for
$\epsilon=0.1$. Analyzing the system performances individually, we note that UW-OFDM
shows a $1$\,dB larger performance loss than CP-OFDM when moving from the idealized
scenario without CFO to $\epsilon=0$. However, UW-OFDM is significantly less
sensitive to an increase of $\epsilon$, as can be seen when comparing the cases for
$\epsilon=0$ and $\epsilon=0.1$. In line with
\cite{Huemer12_1} showing also results without CFO but for a slightly different
system setup (i.e., no pilot subcarriers), UW-OFDM with
$\m{G}''_d$ and CP-OFDM perform very similar for a low coding
rate. While the same still holds true for $\epsilon=0$,
$\m{G}''_d$ offers the same gain of $1.0$\,dB over CP-OFDM for
$\epsilon=0.1$ as $\m{G}'_d$. For $r=3/4$, the
advantages of UW-OFDM due to CFO robustness (both, $\m{G}'_d$ and $\m{G}''_d$) even increase
compared to $r=1/2$, see Fig.~\ref{fig:BER_100ns_cfo_r34_qpsk_g1g2}. As for
$r=1/2$, scenarios without CFO outperform $\epsilon=0$,
but the difference is less prominent and reduces with increasing
$E_b/N_0$, as the impact of the estimation error in $\varphilest$ on the BER
performance becomes less relevant.
\begin{figure}[!tbh]
\centering
\scalebox{0.66}{
%
\definecolor{mycolor1}{rgb}{1.00000,0.50196,0.00000}%
\definecolor{mycolor2}{rgb}{0.12941,0.44314,0.75686}%
\begin{tikzpicture}

\begin{axis}[%
width=4.521in,
height=3.566in,
at={(0.758in,0.481in)},
scale only axis,
unbounded coords=jump,
xmin=0,
xmax=14,
xlabel style={font=\color{white!15!black}},
xlabel={$E_b/N_0$ (dB)},
ymin=-6,
ymax=0,
ylabel style={font=\color{white!15!black}},
ylabel={$\textrm{log}_{10}(\text{BER})$},
axis background/.style={fill=white},
xmajorgrids,
ymajorgrids,
legend style={at={(0.03,0.03)}, anchor=south west, legend cell align=left, align=left, draw=white!15!black}
]
\addplot [color=black, dotted, line width=1.0pt, mark=square*, mark options={solid, fill=black, black}]
  table[row sep=crcr]{%
0	-0.365007372177805\\
1	-0.45379553208721\\
2	-0.6471345600593\\
3	-0.995306721407479\\
4	-1.49215805191861\\
5	-2.07657012082978\\
6	-2.70253959926404\\
7	-3.33760657957974\\
8	-3.97518144810208\\
9	-4.56868655673579\\
10	-5.2641481041684\\
11	-6.06800475478695\\
12	-7.68078861150668\\
};
\addlegendentry{$\m{G}_{d,\cpofdm}$, w/o CFO}

\addplot [color=black, line width=1.0pt, mark=square*, mark options={solid, fill=black, black}]
  table[row sep=crcr]{%
0	-0.324617557783231\\
1	-0.361400154737245\\
2	-0.450676219877569\\
3	-0.641270531899165\\
4	-0.969658861862265\\
5	-1.43683813862612\\
6	-2.01455994495802\\
7	-2.64978433014315\\
8	-3.32873157437155\\
9	-3.99410766404036\\
10	-4.66042732885897\\
11	-5.37651356102955\\
12	-6.17563863318678\\
13	-6.49045691333639\\
14	-7.07872862017872\\
15	-inf\\
};
\addlegendentry{$\m{G}_{d,\cpofdm}$, $\epsilon=0$}

\addplot [color=black, line width=1.0pt, mark=triangle*, mark options={solid, rotate=270, fill=black, black}]
  table[row sep=crcr]{%
0	-0.319520322716873\\
1	-0.347435995308015\\
2	-0.414903477013406\\
3	-0.560043858502742\\
4	-0.818578284871531\\
5	-1.20345337471427\\
6	-1.68853827782393\\
7	-2.23554491327535\\
8	-2.82754683117757\\
9	-3.41945435770755\\
10	-3.9880738506162\\
11	-4.59053055857537\\
12	-5.112\\
13	-5.74425\\
14	-6.39075400014416\\
15	-7.28284860283464\\
};
\addlegendentry{$\m{G}_{d,\cpofdm}$, $\epsilon=0.1$}

\addplot [color=mycolor1, dotted, line width=1.0pt, mark=square*, mark options={solid, fill=mycolor1, mycolor1}]
  table[row sep=crcr]{%
0	-0.422337261214165\\
1	-0.5848053787237\\
2	-0.90303067144592\\
3	-1.3908847807941\\
4	-2.00430371011434\\
5	-2.66972281272628\\
6	-3.36430863288078\\
7	-4.06175597956594\\
8	-4.79140053663878\\
9	-5.63818352341022\\
10	-6.8055008581584\\
11	-inf\\
12	-7.50447086249442\\
};
\addlegendentry{$\m{G}_d'$, w/o CFO}

\addplot [color=mycolor1, line width=1.0pt, mark=square*, mark options={solid, fill=mycolor1, mycolor1}]
  table[row sep=crcr]{%
0	-0.335847219134746\\
1	-0.391707302050244\\
2	-0.51864840439872\\
3	-0.761068195132918\\
4	-1.13884718177929\\
5	-1.63853996838362\\
6	-2.22686164819033\\
7	-2.85964837280339\\
8	-3.47181181645449\\
9	-4.23262925595792\\
10	-4.84076993710477\\
11	-5.57761415354473\\
12	-6.37413709399941\\
13	-6.90241087116646\\
14	-inf\\
15	-inf\\
};
\addlegendentry{$\m{G}_d'$, $\epsilon=0$}

\addplot [color=mycolor1, line width=1.0pt, mark=x, mark options={solid, fill=mycolor1, mycolor1}]
  table[row sep=crcr]{%
0	-0.331209808599357\\
1	-0.37795955692208\\
2	-0.48611995763791\\
3	-0.696681420510446\\
4	-1.03027733389218\\
5	-1.47833141968853\\
6	-2.01287667321306\\
7	-2.613926894836\\
8	-3.25706230057285\\
9	-3.86546162267108\\
10	-4.39153088641034\\
11	-5.06990195846022\\
12	-5.77436363636364\\
13	-6.2909\\
14	-inf\\
15	-inf\\
};
\addlegendentry{$\m{G}_d'$, $\epsilon=0.1$}

\addplot [color=mycolor2, dotted, line width=1.0pt, mark=o, mark options={solid, mycolor2}]
  table[row sep=crcr]{%
0	-0.411226573820478\\
1	-0.544382175269174\\
2	-0.787068090533902\\
3	-1.14886544126949\\
4	-1.60997895722723\\
5	-2.13216287508184\\
6	-2.68498944353647\\
7	-3.28949462777235\\
8	-3.8953034881924\\
9	-4.54878311218091\\
10	-5.29899582575353\\
11	-5.92226\\
12	-6.8055008581584\\
};
\addlegendentry{$\m{G}_d''$, w/o CFO}

\addplot [color=mycolor2, line width=1.0pt, mark=o, mark options={solid, mycolor2}]
  table[row sep=crcr]{%
0	-0.335263281384542\\
1	-0.385248239599782\\
2	-0.496319778965861\\
3	-0.705641081632884\\
4	-1.03069910324371\\
5	-1.46331572240806\\
6	-1.9885030200895\\
7	-2.58458249460185\\
8	-3.2277448854269\\
9	-3.89184592307178\\
10	-4.51769912822817\\
11	-5.365\\
12	-6.358\\
13	-6.57505193678013\\
14	-inf\\
15	-inf\\
};
\addlegendentry{$\m{G}_d''$, $\epsilon=0$}

\addplot [color=mycolor2, line width=1.0pt, mark=triangle, mark options={solid, rotate=90, mycolor2}]
  table[row sep=crcr]{%
0	-0.33048382628098\\
1	-0.374340214374648\\
2	-0.470858933398112\\
3	-0.653534500369031\\
4	-0.943786434602242\\
5	-1.3389033847925\\
6	-1.82492623085921\\
7	-2.37707609113496\\
8	-2.99373153558112\\
9	-3.6188552475624\\
10	-4.32020541943231\\
11	-5.04057787350851\\
12	-5.70792\\
13	-6.72631961211078\\
14	-6.40756084948636\\
15	-inf\\
};
\addlegendentry{$\m{G}_d''$, $\epsilon=0.1$}

\end{axis}
\end{tikzpicture}%
}
\caption{BER results for coded
  transmission with $r=1/2$ and QPSK.}
\label{fig:BER_100ns_cfo_r12_qpsk_g1g2}
\end{figure}
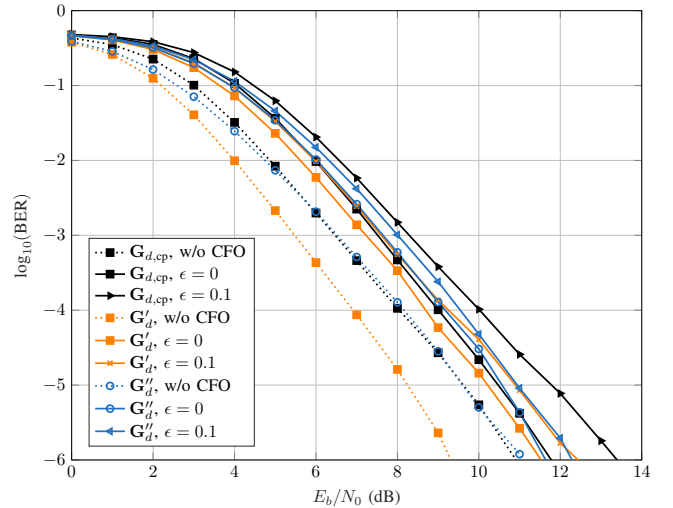
\begin{figure}[!tbh]
\centering
\scalebox{0.66}{
%
\definecolor{mycolor1}{rgb}{1.00000,0.50196,0.00000}%
\definecolor{mycolor2}{rgb}{0.12941,0.44314,0.75686}%
\begin{tikzpicture}

\begin{axis}[%
width=4.521in,
height=3.566in,
at={(0.758in,0.481in)},
scale only axis,
unbounded coords=jump,
xmin=0,
xmax=20,
xlabel style={font=\color{white!15!black}},
xlabel={$E_b/N_0$ (dB)},
ymin=-6,
ymax=0,
ylabel style={font=\color{white!15!black}},
ylabel={$\textrm{log}_{10}(\text{BER})$},
axis background/.style={fill=white},
xmajorgrids,
ymajorgrids,
legend style={at={(0.03,0.03)}, anchor=south west, legend cell align=left, align=left, draw=white!15!black}
]
\addplot [color=black, dotted, line width=1.0pt, mark=square*, mark options={solid, fill=black, black}]
  table[row sep=crcr]{%
0	-0.310155514899433\\
1	-0.32104870840292\\
2	-0.34676900286241\\
3	-0.406606059513155\\
4	-0.532841324879286\\
5	-0.739968859804104\\
6	-1.01459587302043\\
7	-1.33274173743579\\
8	-1.67561714602902\\
9	-2.03806894403289\\
10	-2.41679019420894\\
11	-2.80728380148079\\
12	-3.2181922494771\\
13	-3.64731476799768\\
14	-4.02232405042713\\
15	-4.50931058123859\\
16	-4.9459\\
17	-5.2865\\
18	-5.6375\\
19	-6.124625\\
};
\addlegendentry{$\m{G}_{d,\cpofdm}$, w/o CFO}

\addplot [color=black, line width=1.0pt, mark=square*, mark options={solid, fill=black, black}]
  table[row sep=crcr]{%
0	-0.305470538877384\\
1	-0.311244181125119\\
2	-0.325240897243569\\
3	-0.359825965238233\\
4	-0.441714309516469\\
5	-0.604310477875809\\
6	-0.858871215142785\\
7	-1.18309593059164\\
8	-1.54981746467323\\
9	-1.93798671824202\\
10	-2.33878080538561\\
11	-2.74590948688638\\
12	-3.15846052876619\\
13	-3.59901643486588\\
14	-3.98686107713618\\
15	-4.46937406965248\\
16	-4.86425643401727\\
17	-5.22305764248295\\
18	-5.66135\\
19	-6.1003\\
20	-6.58395\\
};
\addlegendentry{$\m{G}_{d,\cpofdm}$, $\epsilon=0$}

\addplot [color=black, line width=1.0pt, mark=triangle*, mark options={solid, rotate=270, fill=black, black}]
  table[row sep=crcr]{%
0	-0.304486069514527\\
1	-0.308800722343165\\
2	-0.318648084199748\\
3	-0.341849676802511\\
4	-0.395044564413172\\
5	-0.505902346006206\\
6	-0.694692797630642\\
7	-0.958516125044207\\
8	-1.27727806051795\\
9	-1.63072327035844\\
10	-2.00048587267466\\
11	-2.38046869558467\\
12	-2.77318073854008\\
13	-3.15476401482129\\
14	-3.51448968964932\\
15	-3.91363522176367\\
16	-4.25366987392424\\
17	-4.60332941726064\\
18	-4.91328903244131\\
19	-5.16483\\
20	-5.4058\\
};
\addlegendentry{$\m{G}_{d,\cpofdm}$, $\epsilon=0.1$}

\addplot [color=mycolor1, dotted, line width=1.0pt, mark=square*, mark options={solid, fill=mycolor1, mycolor1}]
  table[row sep=crcr]{%
0	-0.318719195625665\\
1	-0.341395033607971\\
2	-0.396664428526205\\
3	-0.51843093235294\\
4	-0.733651350199754\\
5	-1.02766117952841\\
6	-1.37347520821667\\
7	-1.75068659267793\\
8	-2.15446161344989\\
9	-2.57498218567863\\
10	-3.03181906431074\\
11	-3.53549895245254\\
12	-4.13227635516565\\
13	-4.71841927583666\\
14	-5.36272527654392\\
15	-6.04229935455205\\
};
\addlegendentry{$\m{G}_d'$, w/o CFO}

\addplot [color=mycolor1, line width=1.0pt, mark=square*, mark options={solid, fill=mycolor1, mycolor1}]
  table[row sep=crcr]{%
0	-0.307905924403526\\
1	-0.317477254258456\\
2	-0.341536782828508\\
3	-0.402059251719751\\
4	-0.53558519518382\\
5	-0.771656413323865\\
6	-1.10609361229368\\
7	-1.51319173427699\\
8	-1.95091826159682\\
9	-2.42206993612135\\
10	-2.88488080020516\\
11	-3.455996655014\\
12	-4.0091067666236\\
13	-4.68647145883705\\
14	-5.35139073214564\\
15	-6.01333565861673\\
16	-7.20366735678702\\
17	-7.07872862017872\\
18	-7.68078861150668\\
};
\addlegendentry{$\m{G}_d'$, $\epsilon=0$}

\addplot [color=mycolor1, line width=1.0pt, mark=x, mark options={solid, fill=mycolor1, mycolor1}]
  table[row sep=crcr]{%
0	-0.306860986649003\\
1	-0.314790770395993\\
2	-0.334244012662554\\
3	-0.382124996465339\\
4	-0.489022889515237\\
5	-0.687128325226055\\
6	-0.980670981611473\\
7	-1.34975681435227\\
8	-1.77638010140855\\
9	-2.23618412397498\\
10	-2.71156612441956\\
11	-3.21316570908886\\
12	-3.74813730944432\\
13	-4.3474421130823\\
14	-5.03782973209689\\
15	-5.801525\\
16	-6.4559\\
17	-6.62009077115307\\
18	-7.3797586158427\\
};
\addlegendentry{$\m{G}_d'$, $\epsilon=0.1$}

\addplot [color=mycolor2, dotted, line width=1.0pt, mark=o, mark options={solid, mycolor2}]
  table[row sep=crcr]{%
0	-0.319486460459695\\
1	-0.343468606352338\\
2	-0.403265542727898\\
3	-0.535728686078613\\
4	-0.76803796245522\\
5	-1.08984523410219\\
6	-1.47578167439447\\
7	-1.90866318861628\\
8	-2.37046072743778\\
9	-2.86514114260081\\
10	-3.38669731902972\\
11	-3.90481428037731\\
12	-4.55791768864225\\
13	-5.2000632325182\\
14	-5.6952625\\
14.5	-6.1718\\
16	-inf\\
17	-inf\\
};
\addlegendentry{$\m{G}_d''$, w/o CFO}

\addplot [color=mycolor2, line width=1.0pt, mark=o, mark options={solid, mycolor2}]
  table[row sep=crcr]{%
0	-0.308299791709838\\
1	-0.31834398837544\\
2	-0.343889062060138\\
3	-0.408980995068945\\
4	-0.549109153392823\\
5	-0.795944158635125\\
6	-1.14450042457317\\
7	-1.57225724351789\\
8	-2.05274572206172\\
9	-2.58291206670998\\
10	-3.13043793920507\\
11	-3.70948956389664\\
12	-4.31212989911446\\
13	-4.98662231557348\\
14	-5.54884931629626\\
15	-6.25754273756987\\
16	-7.3797586158427\\
17	-inf\\
};
\addlegendentry{$\m{G}_d''$, $\epsilon=0$}

\addplot [color=mycolor2, line width=1.0pt, mark=triangle, mark options={solid, rotate=90, mycolor2}]
  table[row sep=crcr]{%
0	-0.307314818328969\\
1	-0.315564016218747\\
2	-0.335990295318017\\
3	-0.387095080369315\\
4	-0.500834092719101\\
5	-0.705313582506824\\
6	-1.00600047143096\\
7	-1.38993429171166\\
8	-1.83205927262216\\
9	-2.31408835473979\\
10	-2.8211880337078\\
11	-3.37631962665251\\
12	-3.93074329949491\\
13	-4.57154764291848\\
14	-5.145\\
15	-5.858\\
16	-6.319\\
17	-6.49045691333639\\
};
\addlegendentry{$\m{G}_d''$, $\epsilon=0.1$}

\end{axis}
\end{tikzpicture}%
}
\caption{BER results for coded
  transmission with $r=3/4$ and QPSK.}
\label{fig:BER_100ns_cfo_r34_qpsk_g1g2}
\end{figure}
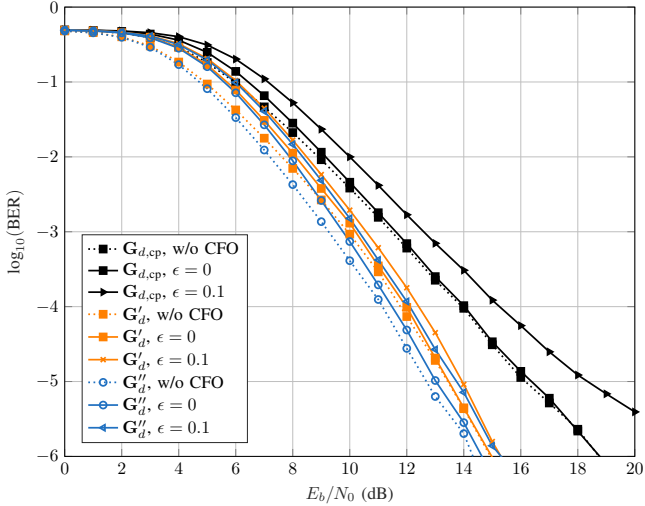
Figs.~\ref{fig:BER_100ns_cfo_r12_qam16_g1g2} and \ref{fig:BER_100ns_cfo_r34_qam16_g1g2}
show the BER performance for $r=1/2$ and $r=3/4$, respectively, when utilizing QAM16
as a representative for higher-order modulation alphabets. While in the low CFO
case (i.e., w/o CFO and $\epsilon=0$), only $\m{G}'_d$ may outperform CP-OFDM,
the superiority of UW-OFDM becomes independent of the specific generator matrix
instance for an increasing CFO, as
illustrated by the case $\epsilon=0.1$ and the thereof resulting saturating BER behavior
of CP-OFDM.
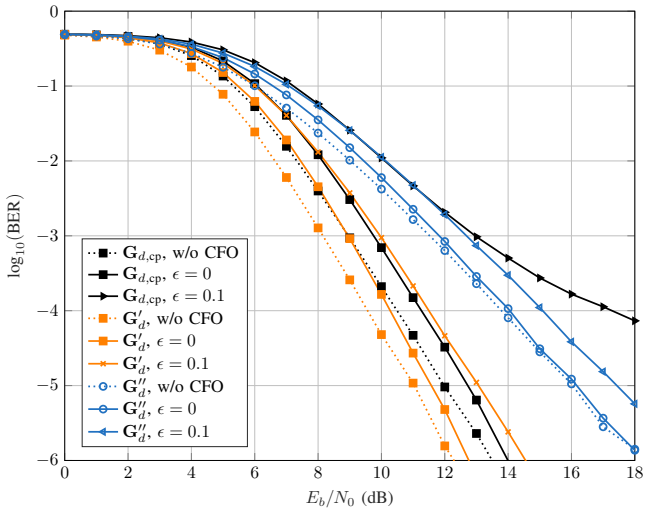
\begin{figure}[!tbh]
\centering
\scalebox{0.66}{
%
\definecolor{mycolor1}{rgb}{1.00000,0.50196,0.00000}%
\definecolor{mycolor2}{rgb}{0.12941,0.44314,0.75686}%
\begin{tikzpicture}

\begin{axis}[%
width=4.521in,
height=3.566in,
at={(0.758in,0.481in)},
scale only axis,
unbounded coords=jump,
xmin=0,
xmax=18,
xlabel style={font=\color{white!15!black}},
xlabel={$E_b/N_0$ (dB)},
ymin=-6,
ymax=0,
ylabel style={font=\color{white!15!black}},
ylabel={$\textrm{log}_{10}(\text{BER})$},
axis background/.style={fill=white},
xmajorgrids,
ymajorgrids,
legend style={at={(0.03,0.03)}, anchor=south west, legend cell align=left, align=left, draw=white!15!black}
]
\addplot [color=black, dotted, line width=1.0pt, mark=square*, mark options={solid, fill=black, black}]
  table[row sep=crcr]{%
0	-0.313617401000391\\
1	-0.329288081270964\\
2	-0.364199001898272\\
3	-0.439981498524761\\
4	-0.593221963277819\\
5	-0.866381173915284\\
6	-1.27545016470153\\
7	-1.80458178153902\\
8	-2.40130812442112\\
9	-3.0263476099391\\
10	-3.67851414700924\\
11	-4.32897725425628\\
12	-5.01849324549739\\
13	-5.63962229824928\\
14	-6.3628\\
};
\addlegendentry{$\m{G}_{d,\cpofdm}$, w/o CFO}

\addplot [color=black, line width=1.0pt, mark=square*, mark options={solid, fill=black, black}]
  table[row sep=crcr]{%
0	-0.308157132606302\\
1	-0.317415623619548\\
2	-0.338188128006655\\
3	-0.38432722357982\\
4	-0.481832861931366\\
5	-0.665834946136138\\
6	-0.968222244863866\\
7	-1.39142185068917\\
8	-1.91745971528575\\
9	-2.51603790629662\\
10	-3.15802007988873\\
11	-3.82561698675344\\
12	-4.48670881975795\\
13	-5.19325\\
14	-6.0045\\
15	-7.13694693905723\\
16	-7.37998498774353\\
};
\addlegendentry{$\m{G}_{d,\cpofdm}$, $\epsilon=0$}

\addplot [color=black, line width=1.0pt, mark=triangle*, mark options={solid, rotate=270, fill=black, black}]
  table[row sep=crcr]{%
0	-0.306409474134717\\
1	-0.312875932582454\\
2	-0.326707150868186\\
3	-0.355183334935587\\
4	-0.412002326266885\\
5	-0.51674030766623\\
6	-0.687744577033298\\
7	-0.93149867207138\\
8	-1.23961232929873\\
9	-1.58885870926857\\
10	-1.9593277929398\\
11	-2.33295021718547\\
12	-2.68564088264827\\
13	-3.013\\
14	-3.29671982563545\\
15	-3.56262623814193\\
16	-3.77861770539531\\
17	-3.9474\\
18	-4.1353\\
19	-4.2461\\
20	-4.3521\\
};
\addlegendentry{$\m{G}_{d,\cpofdm}$, $\epsilon=0.1$}

\addplot [color=mycolor1, dotted, line width=1.0pt, mark=square*, mark options={solid, fill=mycolor1, mycolor1}]
  table[row sep=crcr]{%
0	-0.320113353067266\\
1	-0.345043289186479\\
2	-0.401156193624014\\
3	-0.5204476390042\\
4	-0.746005696510349\\
5	-1.10973484742426\\
6	-1.61304750492406\\
7	-2.22111253346554\\
8	-2.89424886259538\\
9	-3.58819870804135\\
10	-4.31891442119472\\
11	-4.96667686590002\\
12	-5.80584054881467\\
13	-6.47340208889907\\
};
\addlegendentry{$\m{G}_d'$, w/o CFO}

\addplot [color=mycolor1, line width=1.0pt, mark=square*, mark options={solid, fill=mycolor1, mycolor1}]
  table[row sep=crcr]{%
0	-0.310640001893125\\
1	-0.323993555426631\\
2	-0.355029646768414\\
3	-0.424345315361831\\
4	-0.566461527239688\\
5	-0.819405702492837\\
6	-1.20522780969187\\
7	-1.7198911779392\\
8	-2.34495021986849\\
9	-3.03779103040801\\
10	-3.78242345897357\\
11	-4.56641593080724\\
12	-5.31941006996024\\
13	-6.21477594178817\\
14	-6.90275056182273\\
15	-inf\\
16	-inf\\
};
\addlegendentry{$\m{G}_d'$, $\epsilon=0$}

\addplot [color=mycolor1, line width=1.0pt, mark=x, mark options={solid, fill=mycolor1, mycolor1}]
  table[row sep=crcr]{%
0	-0.309104834391941\\
1	-0.320043070687534\\
2	-0.344394120985576\\
3	-0.397688443716662\\
4	-0.505391158509312\\
5	-0.697466670241296\\
6	-0.994016196803961\\
7	-1.39175013313753\\
8	-1.88102209009001\\
9	-2.42685157389837\\
10	-3.02456639489211\\
11	-3.66896150893916\\
12	-4.3349157955147\\
13	-4.95725\\
14	-5.61831982797821\\
15	-6.31102\\
16	-7.62974928975899\\
};
\addlegendentry{$\m{G}_d'$, $\epsilon=0.1$}

\addplot [color=mycolor2, dotted, line width=1.0pt, mark=o, mark options={solid, mycolor2}]
  table[row sep=crcr]{%
0	-0.315776621718392\\
1	-0.333001233195734\\
2	-0.368959420884023\\
3	-0.439469018835998\\
4	-0.562207564528429\\
5	-0.748789852358345\\
6	-0.996262217148326\\
7	-1.29389786058053\\
8	-1.62800678156648\\
9	-1.99055674598126\\
10	-2.37583971039947\\
11	-2.78341423981111\\
12	-3.19745359731498\\
13	-3.63998810204021\\
14	-4.09517325636451\\
15	-4.55008576217163\\
16	-4.9778254845907\\
17	-5.55056804371137\\
18	-5.84919196960947\\
};
\addlegendentry{$\m{G}_d''$, w/o CFO}

\addplot [color=mycolor2, line width=1.0pt, mark=o, mark options={solid, mycolor2}]
  table[row sep=crcr]{%
0	-0.308871435677821\\
1	-0.318868871138021\\
2	-0.340703991340192\\
3	-0.386170083466005\\
4	-0.473754090841176\\
5	-0.620911626701172\\
6	-0.838142911866372\\
7	-1.1188549292117\\
8	-1.45203660685722\\
9	-1.82287522574252\\
10	-2.22201233411559\\
11	-2.64526398274885\\
12	-3.07560350145375\\
13	-3.54246489830523\\
14	-3.97202433555388\\
15	-4.50786130465531\\
16	-4.91318951493737\\
17	-5.43477268654294\\
18	-5.86383249579236\\
};
\addlegendentry{$\m{G}_d''$, $\epsilon=0$}

\addplot [color=mycolor2, line width=1.0pt, mark=triangle, mark options={solid, rotate=90, mycolor2}]
  table[row sep=crcr]{%
0	-0.307922119319367\\
1	-0.316101120075352\\
2	-0.334030326895345\\
3	-0.370663340220773\\
4	-0.440694518220842\\
5	-0.560106903545818\\
6	-0.738591099374413\\
7	-0.977452267902574\\
8	-1.26491286070731\\
9	-1.59144963590747\\
10	-1.94808723195874\\
11	-2.32380063881301\\
12	-2.71630456797011\\
13	-3.13049763606581\\
14	-3.52618500970803\\
15	-3.95780546427103\\
16	-4.41508202007596\\
17	-4.81350798976721\\
18	-5.2435476843582\\
};
\addlegendentry{$\m{G}_d''$, $\epsilon=0.1$}

\end{axis}
\end{tikzpicture}%
}
\caption{BER results coded transmission with
  $r=1/2$ and QAM16.}
\label{fig:BER_100ns_cfo_r12_qam16_g1g2}
\end{figure}
\begin{figure}[!tbh]
\centering
\scalebox{0.66}{
%
\definecolor{mycolor1}{rgb}{1.00000,0.50196,0.00000}%
\definecolor{mycolor2}{rgb}{0.12941,0.44314,0.75686}%
\begin{tikzpicture}

\begin{axis}[%
width=4.521in,
height=3.566in,
at={(0.758in,0.481in)},
scale only axis,
xmin=0,
xmax=21,
xlabel style={font=\color{white!15!black}},
xlabel={$E_b/N_0$ (dB)},
ymin=-6,
ymax=0,
ylabel style={font=\color{white!15!black}},
ylabel={$\textrm{log}_{10}(\text{BER})$},
axis background/.style={fill=white},
xmajorgrids,
ymajorgrids,
legend style={at={(0.03,0.03)}, anchor=south west, legend cell align=left, align=left, draw=white!15!black}
]
\addplot [color=black, dotted, line width=1.0pt, mark=square*, mark options={solid, fill=black, black}]
  table[row sep=crcr]{%
0	-0.302688309935258\\
1	-0.304417722480149\\
2	-0.307833820888975\\
3	-0.314534284707863\\
4	-0.329430747676288\\
5	-0.362328452000071\\
6	-0.434256009052666\\
7	-0.575267292815813\\
8	-0.799687010988231\\
9	-1.095354963131\\
10	-1.4365271165862\\
11	-1.80445599735966\\
12	-2.19379504235699\\
13	-2.60016584320586\\
14	-3.02282456083171\\
15	-3.46384342280905\\
16	-3.89589112476607\\
17	-4.35111657013948\\
18	-4.80796365087994\\
19	-5.1864\\
20	-5.6353\\
21	-6.00508333333333\\
};
\addlegendentry{$\m{G}_{d,\cpofdm}$, w/o CFO}

\addplot [color=black, line width=1.0pt, mark=square*, mark options={solid, fill=black, black}]
  table[row sep=crcr]{%
0	-0.302112790426598\\
1	-0.303249351821293\\
2	-0.305610934708183\\
3	-0.310333969830188\\
4	-0.320356600481644\\
5	-0.342249073459704\\
6	-0.391855228273087\\
7	-0.496081473438721\\
8	-0.681979223966784\\
9	-0.955718420732744\\
10	-1.29911776473085\\
11	-1.68539146956124\\
12	-2.09508518429679\\
13	-2.51756876831879\\
14	-2.95567976142006\\
15	-3.40036411784757\\
16	-3.85165798087684\\
17	-4.31323773106721\\
18	-4.76027166054206\\
19	-5.18585\\
20	-5.5988\\
21	-5.99208333333333\\
};
\addlegendentry{$\m{G}_{d,\cpofdm}$, $\epsilon=0$}

\addplot [color=black, line width=1.0pt, mark=triangle*, mark options={solid, rotate=270, fill=black, black}]
  table[row sep=crcr]{%
0	-0.3018077786516\\
1	-0.302611086422095\\
2	-0.304106497094011\\
3	-0.306817074328998\\
4	-0.311829669934539\\
5	-0.321129877143238\\
6	-0.338482405149663\\
7	-0.370621156075472\\
8	-0.426898357430752\\
9	-0.517027294059569\\
10	-0.643783204543177\\
11	-0.802106593984256\\
12	-0.985431845731041\\
13	-1.18038156054931\\
14	-1.37476734194424\\
15	-1.56257895266875\\
16	-1.73690074317698\\
17	-1.89338865358419\\
18	-2.02861747410135\\
19	-2.14126770355228\\
20	-2.23842802929101\\
21	-2.3148\\
22	-2.3786\\
23	-2.4298\\
24	-2.4752\\
25	-2.5046\\
};
\addlegendentry{$\m{G}_{d,\cpofdm}$, $\epsilon=0.1$}

\addplot [color=mycolor1, dotted, line width=1.0pt, mark=square*, mark options={solid, fill=mycolor1, mycolor1}]
  table[row sep=crcr]{%
0	-0.30355498488028\\
1	-0.306395973791705\\
2	-0.312144643515511\\
3	-0.324264578733287\\
4	-0.351269471992106\\
5	-0.412767660968624\\
6	-0.539123700574534\\
7	-0.754507250750784\\
8	-1.05291549209694\\
9	-1.40756349454923\\
10	-1.79900430398007\\
11	-2.21175865460415\\
12	-2.63640075651921\\
13	-3.08255004602188\\
14	-3.60561402789361\\
15	-4.15238748075344\\
16	-4.67412127545961\\
17	-5.3762014400134\\
18	-6.0045\\
};
\addlegendentry{$\m{G}_d'$, w/o CFO}

\addplot [color=mycolor1, line width=1.0pt, mark=square*, mark options={solid, fill=mycolor1, mycolor1}]
  table[row sep=crcr]{%
0	-0.302492068163098\\
1	-0.304274056398791\\
2	-0.307905554745493\\
3	-0.315637401969116\\
4	-0.332448964922339\\
5	-0.370933831027784\\
6	-0.456266322642281\\
7	-0.621097846674966\\
8	-0.881954316616834\\
9	-1.23087477036999\\
10	-1.63635108662703\\
11	-2.06857525600088\\
12	-2.52174564266067\\
13	-2.99582878054045\\
14	-3.51432590951794\\
15	-4.04445186935077\\
16	-4.71370106512043\\
17	-5.27447480297355\\
18	-5.96085568000155\\
};
\addlegendentry{$\m{G}_d'$, $\epsilon=0$}

\addplot [color=mycolor1, line width=1.0pt, mark=x, mark options={solid, fill=mycolor1, mycolor1}]
  table[row sep=crcr]{%
0	-0.302234930671666\\
1	-0.303717973036442\\
2	-0.306427282764731\\
3	-0.31204327646307\\
4	-0.323308901264616\\
5	-0.346887779571102\\
6	-0.396032944116903\\
7	-0.491594837657067\\
8	-0.652277455820993\\
9	-0.88570478664417\\
10	-1.18378413072295\\
11	-1.52818052079951\\
12	-1.91052546445967\\
13	-2.31146239995238\\
14	-2.74404657564392\\
15	-3.1882580855079\\
16	-3.6628868492045\\
17	-4.12335742855526\\
18	-4.6004\\
19	-5.07775\\
20	-5.45815\\
21	-5.8894\\
22	-6.27496666666667\\
};
\addlegendentry{$\m{G}_d'$, $\epsilon=0.1$}

\addplot [color=mycolor2, dotted, line width=1.0pt, mark=o, mark options={solid, mycolor2}]
  table[row sep=crcr]{%
0	-0.303363656363212\\
1	-0.305735313717199\\
2	-0.310466103460031\\
3	-0.319431786814813\\
4	-0.338413908354709\\
5	-0.37748859351904\\
6	-0.45500232675186\\
7	-0.584676460863757\\
8	-0.76937456380002\\
9	-0.99988733341563\\
10	-1.26695088844703\\
11	-1.56250241965934\\
12	-1.87953375222643\\
13	-2.21762149635256\\
14	-2.55991166731711\\
15	-2.92373665525593\\
16	-3.29469887024387\\
17	-3.67214878132677\\
18	-4.03701977782903\\
19	-4.42852603892252\\
20	-4.90214051140477\\
21	-5.29407035903693\\
};
\addlegendentry{$\m{G}_d''$, w/o CFO}

\addplot [color=mycolor2, line width=1.0pt, mark=o, mark options={solid, mycolor2}]
  table[row sep=crcr]{%
0	-0.302424177718664\\
1	-0.303983915181537\\
2	-0.307060173831635\\
3	-0.313233252539486\\
4	-0.325771929637929\\
5	-0.352219222524961\\
6	-0.407830205688994\\
7	-0.511221325728005\\
8	-0.674257184730969\\
9	-0.894583030653513\\
10	-1.16175870575369\\
11	-1.4607003275933\\
12	-1.78915331967186\\
13	-2.13491763831682\\
14	-2.49256533073747\\
15	-2.86510533252073\\
16	-3.2384725048661\\
17	-3.62823449894692\\
18	-4.04618518103361\\
19	-4.40987066545843\\
20	-4.86660089563525\\
21	-5.28525189398975\\
};
\addlegendentry{$\m{G}_d''$, $\epsilon=0$}

\addplot [color=mycolor2, line width=1.0pt, mark=triangle, mark options={solid, rotate=90, mycolor2}]
  table[row sep=crcr]{%
0	-0.302186553938129\\
1	-0.303548511059371\\
2	-0.305879623531049\\
3	-0.310498942392875\\
4	-0.319452989851552\\
5	-0.337138834328655\\
6	-0.371481845865464\\
7	-0.435634011918875\\
8	-0.542274831550561\\
9	-0.695780099248072\\
10	-0.893936597470063\\
11	-1.12811733308093\\
12	-1.39222529476704\\
13	-1.67292039893723\\
14	-1.97482246884691\\
15	-2.28452794612023\\
16	-2.60437909020612\\
17	-2.93725639376785\\
18	-3.257187708092\\
19	-3.59485988895685\\
20	-3.95260032324934\\
21	-4.27699419201938\\
};
\addlegendentry{$\m{G}_d''$, $\epsilon=0.1$}

\end{axis}
\end{tikzpicture}%
}
\caption{BER results for coded transmission with
  $r=3/4$ and QAM16.}
\label{fig:BER_100ns_cfo_r34_qam16_g1g2}
\end{figure}

In a next step, we will extend BER performance considerations from
simple CPE
correction by $\varphilest$ (see Sec.~\ref{sec:cpe_correction}) to the advanced methods presented in
Sec.~\ref{sec:advanced_cfo_comp}. We limit ourselves to 
results for $\m{G}'_d$ only to enhance clarity in the
figures, knowing though that these sophisticated methods provided the same qualitative
performance gain when we conducted the same experiments with $\m{G}''_d$. Since a
considerable gain on top of CPE correction is only applicable for the high
CFO regime, see Fig.~\ref{fig:mse_cfo_3}, investigations are limited to
$\epsilon = 0.1$. Furthermore, our BER simulations demonstrated an increasing gain
with an increasing order of the modulation alphabet, we thus restrict
elaborations to QAM16. Two advanced methods are
considered on top of CPE correction by $\varphilest$, namely
$\varphi_\pil$ in combination with $\mf{\Lambda}_\text{stat}^{-1}$ serving as a
principle performance bound, as well as
$\est{\varphi}_\pil$ together with $\est{\mf{\Lambda}}_\text{stat}^{H}$
denoting a low complexity implementation thereof, where we obtain $\est{\epsilon}$ from
averaging over 200 single estimates of 200 OFDM symbols. Fig.~\ref{fig:BER_100ns_cfo_r12_qam16_g1_advanced_comp} depicts
for UW-OFDM and CP-OFDM in each case four curves, with two serving as
reference. These two reference curves have already been part of
Fig.~\ref{fig:BER_100ns_cfo_r12_qam16_g1g2} and show the performance for simple CPE
correction in case of $\epsilon=0$ and $\epsilon=0.1$ for a coded
transmission with $r=1/2$. We note that for UW-OFDM the complexity
reduced version with $\est{\varphi}_\pil$ and
$\est{\mf{\Lambda}}_\text{stat}^{H}$ coincides with the performance
bound given by compensating with $\varphi_\pil$ and
$\mf{\Lambda}_\text{stat}^{-1}$. These two methods improve simple CPE compensation by
another $1.4$\,dB and thus reduce the margin to the case of $\epsilon=0$ to
$0.5$\,dB. Improvements are also obtained for CP-OFDM, however, the saturating
BER behavior can
unfortunately not fully be canceled. The benefit of advanced CFO compensation
methods even increases for $r=3/4$. Fig.~\ref{fig:BER_100ns_cfo_r34_qam16_g1_advanced_comp} reveals for
UW-OFDM a $2.2$\,dB gain over CPE compensation, leaving a residual gap of
$0.5$\,dB to the performance in case of $\epsilon=0$. For CP-OFDM, we observe the
same behavior as for $r=1/2$, with the advanced compensation methods again only
partially combating a saturating BER performance. As already stated, please
note that for $\m{G}''_d$ we obtained the same conclusions as for
$\m{G}'_d$, i.e., a significant reduction of the residual gap between
$\epsilon=0$ and $\epsilon=0.1$, and only a minor performance loss due to
approximations in the compensation method.
\begin{figure}[!tbh]
\centering
\scalebox{0.66}{
%
\definecolor{mycolor1}{rgb}{1.00000,0.50196,0.00000}%
\begin{tikzpicture}

\begin{axis}[%
width=4.521in,
height=3.566in,
at={(0.758in,0.481in)},
scale only axis,
unbounded coords=jump,
xmin=0,
xmax=16,
xlabel style={font=\color{white!15!black}},
xlabel={$E_b/N_0$ (dB)},
ymin=-6,
ymax=0,
ylabel style={font=\color{white!15!black}},
ylabel={$\textrm{log}_{10}(\text{BER})$},
axis background/.style={fill=white},
xmajorgrids,
ymajorgrids,
legend style={at={(0.03,0.03)}, anchor=south west, legend cell align=left, align=left, draw=white!15!black}
]
\addplot [color=black, line width=1.0pt, mark=square*, mark options={solid, fill=black, black}]
  table[row sep=crcr]{%
0	-0.308157132606302\\
1	-0.317415623619548\\
2	-0.338188128006655\\
3	-0.38432722357982\\
4	-0.481832861931366\\
5	-0.665834946136138\\
6	-0.968222244863866\\
7	-1.39142185068917\\
8	-1.91745971528575\\
9	-2.51603790629662\\
10	-3.15802007988873\\
11	-3.82561698675344\\
12	-4.48670881975795\\
13	-5.19325\\
14	-6.0045\\
15	-7.13694693905723\\
16	-7.37998498774353\\
};
\addlegendentry{$\m{G}_{d,\cp}$, $\epsilon=0$}

\addplot [color=black, line width=1.0pt, mark=triangle*, mark options={solid, rotate=270, fill=black, black}]
  table[row sep=crcr]{%
0	-0.306409474134717\\
1	-0.312875932582454\\
2	-0.326707150868186\\
3	-0.355183334935587\\
4	-0.412002326266885\\
5	-0.51674030766623\\
6	-0.687744577033298\\
7	-0.93149867207138\\
8	-1.23961232929873\\
9	-1.58885870926857\\
10	-1.9593277929398\\
11	-2.33295021718547\\
12	-2.68564088264827\\
13	-3.013\\
14	-3.29671982563545\\
15	-3.56262623814193\\
16	-3.77861770539531\\
};
\addlegendentry{$\m{G}_{d,\cp}$, $\epsilon=0.1$}

\addplot [color=black, line width=1.0pt, mark=o, mark options={solid, black}]
  table[row sep=crcr]{%
0	-0.307794869492174\\
1	-0.31648568455956\\
2	-0.335862166886287\\
3	-0.378200916564382\\
4	-0.466362406072308\\
5	-0.631126009609731\\
6	-0.894538257770246\\
7	-1.25582814175844\\
8	-1.68834020698604\\
9	-2.15551634896886\\
10	-2.63629985519493\\
11	-3.11132220100212\\
12	-3.53966713222338\\
13	-3.95375\\
14	-4.29518536473773\\
15	-4.61347919371116\\
16	-4.90467707378733\\
};
\addlegendentry{$\m{G}_{d,\cp}$, $\epsilon=0.1$, $\mf{\Lambda}_\text{stat}^{-1}$}

\addplot [color=black, dashed, line width=1.0pt, mark=+, mark options={solid, black}]
  table[row sep=crcr]{%
0	-0.307800197812261\\
1	-0.316489460218781\\
2	-0.335920339944485\\
3	-0.37828534070326\\
4	-0.466417806544549\\
5	-0.631123977001758\\
6	-0.894322393698564\\
7	-1.25511615102832\\
8	-1.68693297222196\\
9	-2.15353717666953\\
10	-2.63370798331544\\
11	-3.10898607927179\\
12	-3.53125718699859\\
13	-3.94575\\
14	-4.28416207037669\\
15	-4.59885817959659\\
16	-4.88478766937807\\
};
\addlegendentry{$\m{G}_{d,\cp}$, $\epsilon=0.1$, $\est{\mf{\Lambda}}_\text{stat}^H$}

\addplot [color=mycolor1, line width=1.0pt, mark=square*, mark options={solid, fill=mycolor1, mycolor1}]
  table[row sep=crcr]{%
0	-0.310640001893125\\
1	-0.323993555426631\\
2	-0.355029646768414\\
3	-0.424345315361831\\
4	-0.566461527239688\\
5	-0.819405702492837\\
6	-1.20522780969187\\
7	-1.7198911779392\\
8	-2.34495021986849\\
9	-3.03779103040801\\
10	-3.78242345897357\\
11	-4.56641593080724\\
12	-5.31941006996024\\
13	-6.21477594178817\\
14	-6.90275056182273\\
15	-inf\\
16	-inf\\
};
\addlegendentry{$\m{G}'_d$, $\epsilon=0$}

\addplot [color=mycolor1, line width=1.0pt, mark=x, mark options={solid, fill=mycolor1, mycolor1}]
  table[row sep=crcr]{%
0	-0.309104834391941\\
1	-0.320043070687534\\
2	-0.344394120985576\\
3	-0.397688443716662\\
4	-0.505391158509312\\
5	-0.697466670241296\\
6	-0.994016196803961\\
7	-1.39175013313753\\
8	-1.88102209009001\\
9	-2.42685157389837\\
10	-3.02456639489211\\
11	-3.66896150893916\\
12	-4.3349157955147\\
13	-4.95725\\
14	-5.61831982797821\\
15	-6.31102\\
16	-7.62974928975899\\
};
\addlegendentry{$\m{G}'_d$, $\epsilon=0.1$}

\addplot [color=mycolor1, line width=1.0pt, mark=o, mark options={solid, mycolor1}]
  table[row sep=crcr]{%
0	-0.310309018287945\\
1	-0.323013804391493\\
2	-0.352314431747491\\
3	-0.417765567409957\\
4	-0.551001903392057\\
5	-0.786934848732559\\
6	-1.14321864425165\\
7	-1.61186017613687\\
8	-2.1690378278751\\
9	-2.79037390975123\\
10	-3.44442438369029\\
11	-4.15003005431942\\
12	-4.90958998635303\\
13	-5.5595\\
14	-6.41525\\
15	-inf\\
16	-inf\\
};
\addlegendentry{$\m{G}'_d$, $\epsilon=0.1$, $\varphi_\pil$, $\mf{\Lambda}_\text{stat}^{-1}$}

\addplot [color=mycolor1, dashed, line width=1.0pt, mark=+, mark options={solid, mycolor1}]
  table[row sep=crcr]{%
0	-0.310319647152305\\
1	-0.323025548911083\\
2	-0.352334111302395\\
3	-0.417796409895444\\
4	-0.551071199822151\\
5	-0.786973685663766\\
6	-1.14340808778768\\
7	-1.61178752552003\\
8	-2.16866614748649\\
9	-2.79020630209348\\
10	-3.44489703891293\\
11	-4.14554438854394\\
12	-4.90711536722518\\
13	-5.55875\\
14	-6.38801139043957\\
15	-inf\\
16	-inf\\
};
\addlegendentry{$\m{G}'_d$, $\epsilon=0.1$, $\est{\varphi}_\pil$, $\est{\mf{\Lambda}}_\text{stat}^H$}

\end{axis}
\end{tikzpicture}%
}
\caption{BER results for coded transmission with
  $r=1/2$ and QAM16.}
\label{fig:BER_100ns_cfo_r12_qam16_g1_advanced_comp}
\end{figure}
\begin{figure}[!tbh]
\centering
\scalebox{0.66}{
%
\definecolor{mycolor1}{rgb}{1.00000,0.50196,0.00000}%
\begin{tikzpicture}

\begin{axis}[%
width=4.521in,
height=3.566in,
at={(0.758in,0.481in)},
scale only axis,
unbounded coords=jump,
xmin=0,
xmax=23,
xlabel style={font=\color{white!15!black}},
xlabel={$E_b/N_0$ (dB)},
ymin=-6,
ymax=0,
ylabel style={font=\color{white!15!black}},
ylabel={$\textrm{log}_{10}(\text{BER})$},
axis background/.style={fill=white},
xmajorgrids,
ymajorgrids,
legend style={at={(0.03,0.03)}, anchor=south west, legend cell align=left, align=left, draw=white!15!black}
]
\addplot [color=black, line width=1.0pt, mark=square*, mark options={solid, fill=black, black}]
  table[row sep=crcr]{%
0	-0.302112790426598\\
1	-0.303249351821293\\
2	-0.305610934708183\\
3	-0.310333969830188\\
4	-0.320356600481644\\
5	-0.342249073459704\\
6	-0.391855228273087\\
7	-0.496081473438721\\
8	-0.681979223966784\\
9	-0.955718420732744\\
10	-1.29911776473085\\
11	-1.68539146956124\\
12	-2.09508518429679\\
13	-2.51756876831879\\
14	-2.95567976142006\\
15	-3.40036411784757\\
16	-3.85165798087684\\
17	-4.31323773106721\\
18	-4.76027166054206\\
19	-5.18585\\
20	-5.5988\\
21	-5.99208333333333\\
};
\addlegendentry{$\m{G}_{d,\cp}$, $\epsilon=0$}

\addplot [color=black, line width=1.0pt, mark=triangle*, mark options={solid, rotate=270, fill=black, black}]
  table[row sep=crcr]{%
0	-0.3018077786516\\
1	-0.302611086422095\\
2	-0.304106497094011\\
3	-0.306817074328998\\
4	-0.311829669934539\\
5	-0.321129877143238\\
6	-0.338482405149663\\
7	-0.370621156075472\\
8	-0.426898357430752\\
9	-0.517027294059569\\
10	-0.643783204543177\\
11	-0.802106593984256\\
12	-0.985431845731041\\
13	-1.18038156054931\\
14	-1.37476734194424\\
15	-1.56257895266875\\
16	-1.73690074317698\\
17	-1.89338865358419\\
18	-2.02861747410135\\
19	-2.14126770355228\\
20	-2.23842802929101\\
21	-2.3148\\
22	-2.3786\\
23	-2.4298\\
24	-2.4752\\
25	-2.5046\\
};
\addlegendentry{$\m{G}_{d,\cp}$, $\epsilon=0.1$}

\addplot [color=black, line width=1.0pt, mark=o, mark options={solid, black}]
  table[row sep=crcr]{%
0	-0.302089236159121\\
1	-0.303143444692116\\
2	-0.305383629236336\\
3	-0.309705537498871\\
4	-0.318645565148497\\
5	-0.337930256969805\\
6	-0.380350179034676\\
7	-0.46635587323225\\
8	-0.61890084497911\\
9	-0.842610862025545\\
10	-1.1240590115509\\
11	-1.44401452367604\\
12	-1.79179991973228\\
13	-2.15247126930109\\
14	-2.52046284126633\\
15	-2.87303143746285\\
16	-3.1972084582538\\
17	-3.51820752268304\\
18	-3.77241777803807\\
19	-3.99027435651352\\
20	-4.1686728658936\\
21	-4.3148\\
22	-4.3803\\
23	-4.4796\\
24	-4.5032\\
25	-4.5324\\
};
\addlegendentry{$\m{G}_{d,\cp}$, $\epsilon=0.1$, $\mf{\Lambda}_\text{stat}^{-1}$}

\addplot [color=black, dashed, line width=1.0pt, mark=+, mark options={solid, black}]
  table[row sep=crcr]{%
0	-0.302089236159121\\
1	-0.303143444692116\\
2	-0.305383629236336\\
3	-0.309705537498871\\
4	-0.318645565148497\\
5	-0.337930256969805\\
6	-0.380350179034676\\
7	-0.46635587323225\\
8	-0.61890084497911\\
9	-0.842610862025545\\
10	-1.1240590115509\\
11	-1.44401452367604\\
12	-1.79179991973228\\
13	-2.15247126930109\\
14	-2.52046284126633\\
15	-2.87303143746285\\
16	-3.1972084582538\\
17	-3.51820752268304\\
18	-3.77241777803807\\
19	-3.99027435651352\\
20	-4.1686728658936\\
21	-4.3148\\
22	-4.3803\\
23	-4.4796\\
24	-4.5032\\
25	-4.5324\\
};
\addlegendentry{$\m{G}_{d,\cp}$, $\epsilon=0.1$, $\est{\mf{\Lambda}}_\text{stat}^H$}

\addplot [color=mycolor1, line width=1.0pt, mark=square*, mark options={solid, fill=mycolor1, mycolor1}]
  table[row sep=crcr]{%
0	-0.302492068163098\\
1	-0.304274056398791\\
2	-0.307905554745493\\
3	-0.315637401969116\\
4	-0.332448964922339\\
5	-0.370933831027784\\
6	-0.456266322642281\\
7	-0.621097846674966\\
8	-0.881954316616834\\
9	-1.23087477036999\\
10	-1.63635108662703\\
11	-2.06857525600088\\
12	-2.52174564266067\\
13	-2.99582878054045\\
14	-3.51432590951794\\
15	-4.04445186935077\\
16	-4.71370106512043\\
17	-5.27447480297355\\
18	-5.96085568000155\\
};
\addlegendentry{$\m{G}'_d$, $\epsilon=0$}

\addplot [color=mycolor1, line width=1.0pt, mark=x, mark options={solid, fill=mycolor1, mycolor1}]
  table[row sep=crcr]{%
0	-0.302234930671666\\
1	-0.303717973036442\\
2	-0.306427282764731\\
3	-0.31204327646307\\
4	-0.323308901264616\\
5	-0.346887779571102\\
6	-0.396032944116903\\
7	-0.491594837657067\\
8	-0.652277455820993\\
9	-0.88570478664417\\
10	-1.18378413072295\\
11	-1.52818052079951\\
12	-1.91052546445967\\
13	-2.31146239995238\\
14	-2.74404657564392\\
15	-3.1882580855079\\
16	-3.6628868492045\\
17	-4.12335742855526\\
18	-4.6004\\
19	-5.07775\\
20	-5.45815\\
21	-5.8894\\
22	-6.27496666666667\\
};
\addlegendentry{$\m{G}'_d$, $\epsilon=0.1$}

\addplot [color=mycolor1, line width=1.0pt, mark=o, mark options={solid, mycolor1}]
  table[row sep=crcr]{%
0	-0.302481981560278\\
1	-0.304164274465743\\
2	-0.307613235406187\\
3	-0.31479320248263\\
4	-0.330342668936933\\
5	-0.365631982880952\\
6	-0.44261828510454\\
7	-0.590465396605363\\
8	-0.82470190700803\\
9	-1.1395138695762\\
10	-1.51185683791528\\
11	-1.92108607403947\\
12	-2.35393408647107\\
13	-2.8056990086278\\
14	-3.29211487505327\\
15	-3.83398244361605\\
16	-4.36947302500631\\
17	-4.95063651481987\\
18	-5.6166\\
19	-6.1074\\
20	-6.568575\\
21	-inf\\
22	-inf\\
};
\addlegendentry{$\m{G}'_d$, $\epsilon=0.1$, $\varphi_\pil$, $\mf{\Lambda}_\text{stat}^{-1}$}

\addplot [color=mycolor1, dashed, line width=1.0pt, mark=+, mark options={solid, fill=mycolor1, mycolor1}]
  table[row sep=crcr]{%
0	-0.302470610002643\\
1	-0.304180842341073\\
2	-0.307635298444005\\
3	-0.31480490132025\\
4	-0.330351293165083\\
5	-0.36565747781183\\
6	-0.442591590842174\\
7	-0.590453771049806\\
8	-0.824715767782898\\
9	-1.13937031942232\\
10	-1.51167543485085\\
11	-1.92089287775693\\
12	-2.35345172322023\\
13	-2.80561510236365\\
14	-3.29264287654195\\
15	-3.83291827932888\\
16	-4.36383993681222\\
17	-4.93672200028483\\
18	-5.6148\\
19	-6.115375\\
20	-6.568575\\
21	-inf\\
22	-inf\\
};
\addlegendentry{$\m{G}'_d$, $\epsilon=0.1$, $\est{\varphi}_\pil$, $\est{\mf{\Lambda}}_\text{stat}^H$}

\end{axis}
\end{tikzpicture}%
}
\caption{BER results for coded transmission with
  $r=3/4$ and QAM16.}
\label{fig:BER_100ns_cfo_r34_qam16_g1_advanced_comp}
\end{figure}

We conclude that the methods presented in this section provide a valuable
performance gain over CPE correction at only moderate additional computational
complexity. Since inverting unitary matrices simplifies to taking the complex conjugate of the matrix elements, the additional overhead of the methods presented is essentially limited to a single matrix-vector multiplication per UW-OFDM symbol.

%
%

\section{Conclusion}
\label{sec:conclusion}
In this work we have shown that UW-OFDM offers a better
CFO robustness than CP-OFDM based on MSE analyses. Moreover, the performance
gap increases along an increasing CFO. Various CFO compensation approaches have been considered, from
simple CPE correction to advanced CFO compensation, and the remaining errors
have been derived analytically.  Furthermore, a system level based assessment in terms of the BER performance of a whole
transceiver chain has been conducted. Uncoded as well as coded transmission in
a frequency selective environment confirmed  the
superiority of UW-OFDM over CP-OFDM with respect to CFO impairments.

\bibliographystyle{./IEEEtran}
\bibliography{./IEEEabrv,uwofdm}

\begin{thebibliography}{10}
\providecommand{\url}[1]{#1}
\csname url@samestyle\endcsname
\providecommand{\newblock}{\relax}
\providecommand{\bibinfo}[2]{#2}
\providecommand{\BIBentrySTDinterwordspacing}{\spaceskip=0pt\relax}
\providecommand{\BIBentryALTinterwordstretchfactor}{4}
\providecommand{\BIBentryALTinterwordspacing}{\spaceskip=\fontdimen2\font plus
\BIBentryALTinterwordstretchfactor\fontdimen3\font minus
  \fontdimen4\font\relax}
\providecommand{\BIBforeignlanguage}[2]{{%
\expandafter\ifx\csname l@#1\endcsname\relax
\typeout{** WARNING: IEEEtran.bst: No hyphenation pattern has been}%
\typeout{** loaded for the language `#1'. Using the pattern for}%
\typeout{** the default language instead.}%
\else
\language=\csname l@#1\endcsname
\fi
#2}}
\providecommand{\BIBdecl}{\relax}
\BIBdecl

\bibitem{Huemer10_1}
M.~Huemer, C.~Hofbauer, and J.~B. Huber, ``{The Potential of Unique Words in
  OFDM},'' in \emph{Proc. 15th Int. OFDM Workshop}, Hamburg, Sep. 2010, pp.
  140--144.

\bibitem{Huemer12_1}
M.~Huemer, C.~Hofbauer, and J.~Huber, ``{Non-Systematic Complex Number RS Coded
  OFDM by Unique Word Prefix},'' \emph{{IEEE} Trans. Signal Process.}, vol.~60,
  no.~1, pp. 285--299, Jan. 2012.

\bibitem{Rajabzadeh13}
M.~Rajabzadeh, H.~Steendam, and H.~Khoshbin, ``{Power Spectrum Characterization
  of Systematic Coded UW-OFDM Systems},'' in \emph{Proc. {IEEE} Veh. Technol.
  Conf. (VTC Fall)}, Las Vegas, NV, USA, Sep. 2013, p.~5.

\bibitem{Rajabzadeh14}
M.~Rajabzadeh, H.~Khoshbin, and H.~Steendam, ``{Sidelobe Suppression for
  Non-Systematic Coded UW-OFDM in Cognitive Radio Networks},'' in \emph{Proc.
  Europ. Wireless Conf.}, Barcelona, Spain, May 2014, pp. 826--831.

\bibitem{Rajabzadeh18}
M.~Rajabzadeh and H.~Steendam, ``{Power Spectral Analysis of UW-OFDM
  Systems},'' \emph{{IEEE} Trans. Commun.}, vol.~66, no.~6, pp. 2685--2695,
  Jun. 2018.

\bibitem{Huemer11_1}
M.~Huemer, A.~Onic, and C.~Hofbauer, ``{Classical and Bayesian Linear Data
  Estimators for Unique Word OFDM},'' \emph{{IEEE} Trans. Signal Process.},
  vol.~59, no.~12, pp. 6073--6085, Dec. 2011.

\bibitem{Hofbauer16_1}
C.~Hofbauer, C.~B\"ock, and M.~Huemer, ``{From Dedicated Redundant Subcarriers
  to Distributed Redundancy in UW-OFDM},'' in \emph{Proc. Asilomar Conf.
  Signals, Systems and Computers}, Nov. 2016, pp. 1099--1103.

\bibitem{Qasem2021}
Z.~A.~H. Qasem, J.~Wang, X.~Kuai, H.~Sun, and H.~Esmaiel, ``{Enabling Unique
  Word OFDM for Underwater Acoustic Communication},'' \emph{{IEEE} Wireless
  Commun. Lett.}, vol.~10, no.~9, pp. 1886--1889, 2021.

\bibitem{Chema2016}
S.~A. Cheema, J.~Zhang, M.~Huemer, and M.~Haardt, ``{Linear detection schemes
  for MIMO UW-OFDM},'' in \emph{Proc. Asilomar Conf. Signals, Systems and
  Computers}, 2016, pp. 1457--1461.

\bibitem{Huemer12_2}
M.~Huemer, C.~Hofbauer, A.~Onic, and J.~B. Huber, ``{On the Exploitation of the
  Redundant Energy in UW-OFDM: LMMSE Versus Sphere Detection},'' \emph{{IEEE}
  Signal Process. Lett.}, vol.~19, no.~6, pp. 340--343, Jun. 2012.

\bibitem{Onic14}
A.~Onic and M.~Huemer, ``{Noise Interpolation for Unique Word OFDM},''
  \emph{{IEEE} Signal Process. Lett.}, vol.~21, no.~7, pp. 814--818, Jul. 2014.

\bibitem{Onic13}
A.~Onic, ``{Receiver Concepts for Unique Word OFDM},'' Ph.D. dissertation,
  Institute of Networked and Embedded Systems, Alpen-Adria-Universit\"at
  Klagenfurt, Nov. 2013.

\bibitem{Steendam16}
H.~Steendam, ``{Theoretical Performance Evaluation and Optimization of
  UW-OFDM},'' \emph{{IEEE} Trans. Commun.}, vol.~64, no.~4, pp. 1739--1750,
  Apr. 2016.

\bibitem{Douillard95}
C.~Douillard, M.~Jezequel, C.~Berrou, A.~Picar, P.~Didier, and A.~Glavieux,
  ``{Iterative Correction of Intersymbol Interference: Turbo Equalization},''
  \emph{Proc. Eur. Trans. Telecommun. (ETT)}, vol.~6, no.~3, pp. 507--511, Sep.
  2012.

\bibitem{Tuechler02_1}
M.~{Tuechler}, A.~C. {Singer}, and R.~{Koetter}, ``{Minimum Mean Squared Error
  Equalization using a Priori Information},'' \emph{{IEEE} Trans. Signal
  Process.}, vol.~50, no.~3, pp. 673--683, Mar. 2002.

\bibitem{Haselmayr14}
W.~Haselmayr, C.~Hofbauer, B.~Etzlinger, A.~Springer, and M.~Huemer,
  ``{Iterative Detection for Unique Word OFDM},'' in \emph{Proc. Conf. Global
  Commun. (Globecom)}, Austin, TX, USA, Dec. 2014, pp. 3261--3266.

\bibitem{Haselmayr19}
W.~Haselmayr, C.~Hofbauer, M.~Huemer, and A.~Springer, ``{Approaching the
  Matched Filter Bound with Unique Word OFDM},'' in \emph{Proc. {IEEE} Int.
  Conf. Commun. (ICC)}, May 2019, pp. 1--4.

\bibitem{Baumgartner2023}
S.~Baumgartner, G.~Bognár, O.~Lang, and M.~Huemer, ``{Neural Network
  Approaches for Data Estimation in Unique Word OFDM Systems},'' \emph{{IEEE}
  Trans. Veh. Technol.}, pp. 1--16, 2023.

\bibitem{Bognar2021}
G.~Bognár, S.~Baumgartner, O.~Lang, and M.~Huemer, ``{Neural Network Optimal
  UW-OFDM},'' in \emph{Proc. Asilomar Conf. Signals, Systems and Computers},
  2021, pp. 389--394.

\bibitem{Muck06}
M.~Muck, M.~de~Courville, and P.~Duhamel, ``{A Pseudorandom Postfix OFDM
  Modulator---Semi-Blind Channel Estimation and Equalization},'' \emph{{IEEE}
  Trans. Signal Process.}, vol.~54, no.~3, pp. 1005--1017, Mar. 2006.

\bibitem{Welden08}
D.~V. Welden, H.~Steendam, and M.~Moeneclaey, ``{Iterative DA/DD Channel
  Estimation for KSP-OFDM},'' in \emph{Proc. {IEEE} Int. Conf. Commun. (ICC)},
  Beijing, China, May 2008, pp. 693--697.

\bibitem{Tang07}
S.~Tang, F.~Yang, K.~Peng, C.~Pan, K.~Gong, and Z.~Yang, ``{Iterative Channel
  Estimation for Block Transmission with Known Symbol Padding --- A New Look at
  TDS-OFDM},'' in \emph{Proc. Conf. Global Commun. (Globecom)}, Washington, DC,
  USA, Nov. 2007, pp. 4269--4273.

\bibitem{China06}
``{Framing Structure, Channel Coding and Modulation for Digital Television
  Terrestrial Broadcasting System},'' Chinese National Standard, Std.\ GB 20
  600-2006, 2006.

\bibitem{Ong10}
C.~yen Ong, J.~Song, C.~Pan, and Y.~Li, ``{Technology and Standards of Digital
  Television Terrestrial Multimedia Broadcasting [Topics in Wireless
  Communications]},'' \emph{{IEEE} Commun. Mag.}, vol.~48, no.~5, pp. 119--127,
  May 2010.

\bibitem{Jingyi02}
L.~Jingyi, W.~Hai, P.~Joo, and J.~Ro, ``{The Effect of Filling Unique Words to
  Guard Interval for OFDM System},'' C802.16a-02/87, IEEE 802.16 Broadband
  Wireless Access Working Group, Sep. 2002.

\bibitem{Rajabzadeh2021}
M.~Rajabzadeh and H.~Steendam, ``{Precoding for PAPR Reduction in UW-OFDM},''
  \emph{{IEEE} Commun. Lett.}, vol.~25, no.~7, pp. 2305--2308, 2021.

\bibitem{Huber12_1}
J.~B. Huber, J.~Rettelbach, M.~Seidl, and M.~Huemer, ``{Signal Shaping for
  Unique-Word OFDM by Selected Mapping},'' in \emph{Proc. Europ. Wireless
  Conf.}, Poznan, Poland, Apr. 2012, p.~8.

\bibitem{Rettelbach12}
J.~Rettelbach and J.~B. Huber, ``{PMR-Reduction for Continuous Time OFDM
  Transmit Signals by Selected Mapping},'' in \emph{Proc. Int. Symp. Signals,
  Syst. and Electron. (ISSE)}, Potsdam, Germany, Oct. 2012.

\bibitem{Onic11}
A.~Onic and M.~Huemer, ``{Limiting the Complexity of Sphere Decoding for
  UW-OFDM},'' in \emph{Proc. Int. OFDM Workshop}, Hamburg, Sep. 2011, pp.
  135--139.

\bibitem{IEEE99}
``{IEEE Std 802.11a-1999, Part 11: Wireless LAN Medium Access Control (MAC) and
  Physical Layer (PHY) specifications: High-Speed Physical Layer in the 5 GHz
  Band},'' IEEE, 1999.

\bibitem{Imec00}
L.~Deneire, B.~Gyselinckx, and M.~Engels, ``{Training Sequence versus Cyclic
  Prefix---A New Look on Single Carrier Communication},'' \emph{{IEEE} Commun.
  Lett.}, vol.~5, no.~7, pp. 292--294, Jul. 2001.

\bibitem{Witschnig03}
H.~{Witschnig}, T.~{Mayer}, M.~{Petit}, H.~{Hutzelmann}, A.~{Springer}, and
  R.~{Weigel}, ``{The Advantages of a Unique Word for Synchronisation and
  Channel Estimation in a SC/FDE System},'' in \emph{Proc. Europ. Mobile Comm.
  Conf.}, Apr. 2003, pp. 436--440.

\bibitem{Aboltins2012}
A.~{Aboltins}, ``{Carrier Frequency Offset Estimator based on Unique Word
  Cross-Correlation},'' in \emph{Proc. Telecomm. Forum (TELFOR)}, Nov. 2012,
  pp. 486--489.

\bibitem{Kim2010}
K.~{Kim} and H.~{Park}, ``{Enhanced Phase Tracking for Unique Word based SC-FDE
  on Frequency Selective Channels},'' in \emph{Proc. Int. Microw. Workshop
  Series RF Front-ends Softw. Def. and Cogn. Radio Sol. (IMWS)}, Apr. 2010,
  p.~4.

\bibitem{Ehsa202004}
S.~Ehsanfar, M.~Chafii, and G.~Fettweis, ``{A Study on Unique-Word based
  Synchronization for MIMO Systems over Time-Varying Channels},'' in
  \emph{Proc. Wireless Comm. Netw. Conf. (WCNC)}, Seoul, Korea, Apr. 2020, pp.
  1--7.

\bibitem{Hofbauer20}
C.~Hofbauer, W.~Haselmayr, H.-P. Bernhard, and M.~Huemer, ``{On the Inclusion
  and Utilization of Pilot Tones in Unique Word OFDM},'' \emph{{IEEE} Trans.
  Signal Process.}, vol.~68, pp. 5504--5518, 2020.

\bibitem{Hofbauer20_1}
C.~Hofbauer, W.~Haselmayr, and M.~Huemer, ``{Pilot Tone Insertion and
  Utilization in Unique Word OFDM},'' in \emph{Proc. Int. Workshop Signal Proc.
  Adv. Wireless Comm. (SPAWC)}, Atlanta, GA, USA, May 2020, p.~5.

\bibitem{Hofbauer16}
C.~Hofbauer, ``{Design and Analysis of Unique Word OFDM},'' Ph.D. dissertation,
  Institute of Networked and Embedded Systems, Alpen-Adria-Universit\"at
  Klagenfurt, Jun. 2016.

\bibitem{Hofbauer20_2}
C.~Hofbauer, W.~Haselmayr, H.-P. Bernhard, and M.~Huemer, ``{Impact of a
  Carrier Frequency Offset on Unique Word OFDM},'' in \emph{Proc. Int. Symp.
  Pers., Indoor and Mobile Radio Comm. (PIMRC)}, London, UK, Sep. 2020, p.~7.

\bibitem{Huemer14}
M.~Huemer, C.~Hofbauer, A.~Onic, and J.~B. Huber, ``{Design and Analysis of
  UW-OFDM Signals},'' \emph{Int. J. Electron. and Commun. AEU}, vol.~68,
  no.~10, pp. 958--968, Oct. 2014.

\bibitem{Kay93}
S.~M. Kay, \emph{{Fundamentals of Statistical Signal Processing, Volume I:
  Estimation Theory}}, 1st~ed.\hskip 1em plus 0.5em minus 0.4em\relax Prentice
  Hall, Apr. 1993.

\bibitem{Golomb1965}
S.~W. Golomb and R.~A. Scholtz, ``{Generalized Barker Sequences},''
  \emph{{IEEE} Trans. Inf. Theory}, vol.~11, no.~4, pp. 533--537, Oct. 1965.

\bibitem{Neshaastegaran19}
P.~Neshaastegaran and A.~H. Banihashemi, ``{Log-Likelihood Ratio Calculation
  for Pilot Symbol Assisted Coded Modulation Schemes With Residual Phase
  Noise},'' \emph{{IEEE} Trans. Commun.}, vol.~67, no.~5, pp. 3782--3790, 2019.

\bibitem{Norifumi13}
N.~Kamiya and E.~Sasaki, ``{Pilot-Symbol Assisted and Code-Aided Phase Error
  Estimation for High-Order QAM Transmission},'' \emph{{IEEE} Trans. Commun.},
  vol.~61, no.~10, pp. 4369--4380, 2013.

\bibitem{Cai04}
X.~Cai and G.~B. Giannakis, ``{Error Probability Minimizing Pilots for OFDM
  with M-PSK Modulation over Rayleigh-Fading Channels},'' \emph{{IEEE} Trans.
  Veh. Technol.}, vol.~53, no.~1, pp. 146--155, Jan. 2004.

\bibitem{Fak97}
J.~Fakatselis, \emph{{Criteria for 2.4 GHz PHY Comparison of Modulation}}, IEEE
  Document, 1997, p802.11-97/157r1.

\end{thebibliography}

\end{document}